\documentclass[sigconf]{acmart}
\usepackage{graphicx} 
\usepackage{caption}
\usepackage{subfigure}
\usepackage{balance}

\AtBeginDocument{%
  \providecommand\BibTeX{{%
    \normalfont B\kern-0.5em{\scshape i\kern-0.25em b}\kern-0.8em\TeX}}}

\copyrightyear{2022}
\acmYear{2022}
\setcopyright{acmcopyright}
\acmConference[CIKM '22] {Proceedings of the 31st ACM International Conference on Information and Knowledge Management}{October 17--21, 2022}{Atlanta, GA, USA.}
\acmBooktitle{Proceedings of the 31st ACM International Conference on Information and Knowledge Management (CIKM '22), October 17--21, 2022, Atlanta, GA, USA}
\acmPrice{15.00}
\acmISBN{978-1-4503-9236-5/22/10}
\acmDOI{10.1145/3511808.3557483}

\begin{document}

\title{Unbiased Learning to Rank with Biased Continuous Feedback}


\author{Yi Ren}
\authornote{Both authors contributed equally to this research.}
\affiliation{%
  \institution{Tencent }
  \city{Beijing}
  \country{China}
  \postcode{78229}}
\email{henrybjren@tencent.com}

\author{Hongyan Tang}
\authornotemark[1]
\affiliation{%
  \institution{ Tencent }
  \city{Beijing}
  \country{China}}
\email{violatang@tencent.com}

\author{Siwen Zhu}
\affiliation{%
  \institution{Tencent }
  \city{Beijing}
  \country{China}}
\email{siwenzhu@tencent.com}

\renewcommand{\shortauthors}{Yi Ren, Hongyan Tang, \& Siwen Zhu}

\begin{abstract}
It is a well-known challenge to learn an unbiased ranker with biased feedback. Unbiased learning-to-rank(LTR) algorithms, which are verified to model the relative relevance accurately based on noisy feedback, are appealing candidates and have already been applied in many applications with single categorical labels, such as user click signals. Nevertheless, the existing unbiased LTR methods cannot properly handle continuous feedback, which are essential for many industrial applications, such as content recommender systems. 

To provide personalized high-quality recommendation results, recommender systems need model both categorical and continuous biased feedback, such as click and dwell time. As unbiased LTR methods could not handle these continuous feedback and pair-wise learning without debiasing often performs worse than point-wise on biased feedback, which is also verified in our experiments, training multiple point-wise rankers to predict the absolute value of multiple objectives and leveraging a distinct shallow tower to estimate and alleviate the impact of position bias has been the mainstream approach in major industrial recommendation applications. However, with such a training paradigm, the optimization target differs a lot from the ranking metrics valuing the relative order of top-ranked items rather than the prediction precision of each item. Moreover, as the existing system tends to recommend more relevant items at higher positions, it is difficult for the shallow tower based methods to precisely attribute the user feedback to the impact of position or relevance. Therefore, there exists an exciting opportunity for us to get enhanced performance if we manage to solve the aforementioned issues. 

Accordingly, we design a novel unbiased LTR algorithm \footnote{The source code is released at https://github.com/phyllist/ULTRA.} to tackle the challenges, which innovatively models position bias in the pairwise fashion and introduces the pairwise trust bias to separate the position bias, trust bias, and user relevance explicitly and can work for both continuous and categorical feedback.  Experiment results on public benchmark datasets and internal live traffic of a large-scale recommender system at Tencent News show superior results for continuous labels and also competitive performance for categorical labels of the proposed method.
\end{abstract}





\maketitle

\section{Introduction}

Currently, we see widespread adoption of unbiased learning to rank algorithms \cite{ai2018unbiased,hu2019unbiased,agarwal2019addressing} for many scenarios with biased click signals, such as search engines. These algorithms model position bias as a counterfactual effect and estimate the unbiased relevance preference by weighting each noisy click signal with its Inverse Propensity Weight to get exciting performance gains for the corresponding applications. Nonetheless, the existing methods can only work for categorical labels instead of continuous labels, as the relationship between relevance and continuous user feedback does not adhere to their core assumptions that the binary click probability equals to the multiplication of the binary probability of relevance and examination shown in Equation \ref{eq:3}. Moreover, as continuous feedback (e.g., dwell time and video watch ratio) is essential to accurately model user preference in many other scenarios, such as content recommender systems, it is desirable to design an unbiased LTR method for continuous labels.

Recommender systems need predict both categorical and continuous user feedback (e.g., click, purchase, share, and dwell-time) to learn user preference and recommend top-ranked items to the corresponding user. And user behaviors are often biased in recommendation scenarios. For instance, a user might click and watch a video simply because it is ranked high rather than its relevance. Thus, to estimate relative user preference accurately and improve ranking performance in recommender systems, it is essential to correct the position bias \cite{chen2020bias} from biased user behaviors. As the existing unbiased LTR methods cannot handle continuous feedback appropriately and pair-wise learning without debiasing often performs worse than point-wise on biased feedback, which is also verified by our experiments in Section \ref{section:5}, the mainstream industrial ranking modules usually train multiple point-wise rankers to learn the absolute probability of multiple user behaviors \cite{zhao2019recommending,tang2020progressive} and add a shallow tower to model and eliminate position bias directly \cite{zhao2019recommending,guo2019pal}. However, there is a gap between the training objective of point-wise learning and the actual ranking objective in recommendation scenarios as ranking cares about items' relative order rather than the absolute value. Moreover, bias features and relevance-related features are coupled together with the shallow tower based models, which makes it rather difficult to eliminate bias effectively \cite{wang2018position}. Therefore, it is difficult to correct position bias and learn user preference effectively through such a learning paradigm.

To achieve effective bias correction and incorporate pair-wise learning paradigm, we propose a novel unbiased pair-wise LTR method, which models position-based examination bias in the pairwise fashion and introduces the pairwise trust bias to better model the underlying bias and learn the relative preference more accurately. Compared with the SOTA unbiased LTR methods, our method models the correlation between the unbiased relevance pairs and the biased user feedback pairs directly, which bears no assumption mentioned above on labels and can work for both categorical and continuous labels. Moreover, pairwise debiasing is more consistent with the objective of ranking, which achieves better performance of preference learning. Furthermore, the introduction of pairwise trust bias separates the position-dependent examination bias and the position-pair-dependent relative trust bias explicitly for better correction. Offline experiment on LTR and recommendation benchmark datasets and online A/B testing in a large-scale video recommender system at Tencent News show that our method not only outperforms the SOTA ranking models for continuous labels but also performs competitively for categorical labels.

The main contributions of the paper are as follows:

\begin{itemize}
\item We propose a novel unbiased pairwise LTR method for better bias correction and unbiased relative preference learning. To the best of our knowledge, the proposed method is the first unbiased LTR method that can work for both categorical labels and continuous labels.
\item We extend the regression-based EM algorithm \cite{wang2018position} to the pairwise setting to estimate the parameters of the proposed method effectively.

\item We conduct offline and online experiments to evaluate the performance of our method. Results on benchmark datasets show that our method achieves competitive performance for both continuous and categorical labels. Online A/B testing in a large-scale recommender system at Tencent News shows a significant 2.08\% improvement of the business metric.

\end{itemize}

\section{Related Work}
In this section, we discuss related works on unbiased learning to rank and debiasing methods in recommender systems.
\subsection{Unbiased Learning to Rank Algorithms}
Recently, unbiased LTR has been actively studied as a promising approach to learning from biased feedback for position bias \cite{chen2020bias} correction. Joachims \cite{joachims2017unbiased} first present the counterfactual framework to learn the theoretically unbiased ranker via the inverse propensity weighting (IPW) estimator. To estimate the propensity, both Wang \cite{wang2016learning} and Joachims \cite{joachims2017unbiased} propose the methods of result randomization and intervention. To reduce the negative impact on user experience, some works propose to learn the propensity from biased feedback directly, such as DLA \cite{ai2018unbiased}, regression EM \cite{wang2018position}, and Unbiased LambdaMART \cite{hu2019unbiased}. The position-dependent click noise \cite{joachims2017accurately} is ignored in these methods, which is addressed by \cite{agarwal2019addressing}. However, all of the above methods can only work for categorical labels due to their core assumption on binary click probability and binary relevance probability shown in equation \ref{eq:3}. In contrast, our proposed method can work for and achieves competitive performance for both categorical and continuous labels, according to experiment results in this paper.

\subsection{Unbiased Recommender Learning from Missing-Not-At-Random Implicit Feedback}
One related but different line of research \cite{hu2008collaborative,liang2016modeling,saito2020unbiased,saito2020unbiased2,lee2021dual} focuses on learning unbiased recommenders based on the Missing-Not-At-Random (MNAR) implicit feedback. These works mainly focus on eliminating exposure bias  \cite{chen2020bias} and selection bias due to MNAR feedback  \cite{chen2020bias} for collaborative filtering based recommendation methods \cite{koren2009matrix,rendle2012bpr,johnson2014logistic}. Collaborative filtering works well for the candidate generation module in recommender systems to retrieve a set of relevant items from all candidates in low latency \cite{huang2020embedding}, which considers interacted items as positive and all the other items as potential negative samples but often ignores the hard negative samples. In contrast, our work targets the ranking module and adopts hard negative items that are impressed but not interacted as negative samples to rank the most desired items on the top. Moreover, our work focuses on correcting the position bias \cite{chen2020bias} rather than the selection bias and exposure bias.



\subsection{Modeling Position Bias for the Ranking Module of the Recommender Systems}
As mentioned above, the ranking module accepts the impressed but not interacted items as negative samples for accurate recommendation. Conventional ranking modules often adopt pointwise learning methods to estimate the interaction probability \cite{zhao2019recommending,tang2020progressive} as unbiased LTR methods could not deal with biased continuous feedback. To correct position bias \cite{chen2020bias} for  pointwise ranking models,  Zhao \cite{zhao2019recommending} and Guo \cite{guo2019pal} both propose to employ a shallow tower to estimate and alleviate bias by combining the output of the shallow tower and the main model with summation or multiplication. However, due to the coupling of bias features and relevance-related features, shallow tower based methods cannot work effectively \cite{wang2018position}. Wu \cite{wu2021unbiased} applies the idea of \cite{wang2018position} to learn unbiased ranker with IPW framework and regression-EM for recommendation. However, it still can only work for categorical labels, which is inadequate for typical recommender systems. Our method extends the IPW framework to the pairwise setting to deal with continuous labels, which achieves superior performance of debiasing and preference learning.

\section{Methodology }
In this section, we provide the general framework of unbiased learning to rank for both categorical and continuous labels. 
\subsection{Unbiased LTR for Categorical Labels}
Suppose there are $N$ positions, for a user request $u$, item $x_i$ is displayed at position $i  \in  [1, N]$, $r_i$ is the unbiased relevance of  user-item pair $(u,x_i)$. For simplicity, we only consider binary relevance, and one can easily extend it to the multi-level case. Then we can calculate the risk function as follows:
\begin{equation}
R_{rel}(f) = \int_{}{}L(f(u,x_i), r_i) dP(r_i=1,u,x_i)
\end{equation}
where $f(u,x_i)$ denotes the ranking score for user-item pair $(u,x_i)$, $L(f(u,x_i),r_i)$ denotes the loss function based on performance metrics and $P(r_i \! \! = \!\! 1,u,x_i)$ denotes the probability distribution of positive $r_i$ on $x_i$. For simplicity, the position information of items is omitted from the loss function. As most performance metrics such as Discounted Cumulative Gain (DCG) \cite{jarvelin2002cumulated,jarvelin2017ir} and Average Relevance Position (ARP) \cite{wang2018lambdaloss} only consider relevant items, the risk function here is calculated only based on items with positive relevance label.


\begin{equation}
\hat{f}_{rel} = \arg\min_{f}  \sum_{u,x_i,r_i>0}{} L(f(u,x_i), r_i) 
\end{equation}


If the unbiased relevance label is available, we can learn a ranker by minimizing the empirical risk function shown in Equation 2. However, it is infeasible to collect enough true relevance labels in large-scale applications as we cannot afford to find enough suitable judges for all users. Thus, abundant biased user actions (e.g., click, purchase, like) in real-world applications are valuable labels if we can fill the gap between user actions and relevance labels. As assumed in Position-Based Model (PBM) \cite{richardson2007predicting}, the user clicks the item if and only if he examines a relevant item, and the probability of examination only depends on the position and is independent of relevance. As shown in \autoref{table0}, let $e_i$ denotes whether item $x_i$ is examined by the user, $c_i$ denotes the user action label of $x_i$, we can model the relationship between $c_i$ and $r_i$ as follows:
\begin{align} \label{eq:3}
P(c_i=1|u,x_i,i)&=P(r_i=1|u,x_i,i) P (e_i=1|u,x_i,i) \nonumber  \\
&= P(r_i=1|u,x_i) P (e_i=1|i)
\end{align}

Then we can derive the risk function and empirical risk function based on the user action label:
\begin{align} \label{eq:4}
R_{unbiased}(f) &= \int_{}{}  \frac{L(f(u,x_i), c_i)} {P (e_i=1|i)} dP(c_i=1,u,x_i,i)  \vspace{1ex} \nonumber \\
& = \int_{}{}  L(f(u,x_i), c_i) d \frac {P(c_i=1,u,x_i,i) } {P (e_i=1|i)} \nonumber \\
& = \int_{}{}  L(f(u,x_i), c_i)  dP(r_i=1,u,x_i) \nonumber \\
& = \int_{}{}  L(f(u,x_i), r_i)  dP(r_i=1,u,x_i) \nonumber \\
& = R_{rel}(f)
\end{align}

\vspace{-0.4cm}
\begin{equation} \label{eq:5}
\hat{f}_{rel} = \hat{f}_{unbiased} = \arg\min_{f}  \sum_{u,x_i,i,c_i>0}{} \frac {L(f(u,x_i), c_i)} {P (e_i=1|i)}
\end{equation}

As shown in Equation \ref{eq:4}, $R_{unbiased}$ with the inverse propensity weighting (IPW) \cite{joachims2017unbiased} loss equals to $R_{rel}$ in fact. In other words, if we can estimate the examination probability $P(e_i=1|i)$ accurately, we can learn an unbiased ranker based on user action labels.

Due to the assumption between the probability distribution of relevance labels and user action labels shown in Equation \ref{eq:3}, the aforementioned method can only be applied to tasks with categorical labels rather than continuous labels, which tends to exhibit much more complex relationships between the user action labels and the true relevance. However, continuous labels such as dwell time and video watch ratio are important feedback of user satisfaction and are commonly used in industrial recommender systems. To correct position bias for continuous labels, we propose a novel unbiased LTR method which circumvents the previous assumption between unbiased relevance labels and biased user action labels. Details of the method would be provided in the following sections.

\begin{table}
\setlength{\belowcaptionskip}{0.1cm}
\begin{tabular}{cl}
\hline
Notation & Description \\
\hline
$u$ & User request with user profile and context. \\
$x_i$ & Features of item displayed in position $i$. \\
$r_i$ & Unbiased relevance of user-item pair $(u,x_i)$. \\
$c_i$ & Biased feedback label of user-item pair $(u,x_i)$. \\
$e_i$ & Examination of $x_i$. \\
$\theta_i$ & $P (e_i=1|i)$, the probability of examination. \\
$\theta_i^-$ & $P (e_i=1|i,c_i=0)$ \\
$\epsilon_{ij}^+$ & $P(c_i>c_j | e_i=1,e_j=1,r_i>r_j,i,j)$ \\
$\epsilon_{ij}^-$ & $P(c_i>c_j | e_i=1,e_j=1,r_i<=r_j,i,j)$ \\
$ \gamma_{u,x_i,x_j}$ & $P(r_i>r_j|u,x_i,x_j)$ \\
$\beta_{u,x_i}$ & $P(r_i>0|u,x_i)$, the relevant probability of $(u,x_i)$. \\
\hline
\end{tabular}
\caption{Notations and Descriptions}
\label{table0}
\end{table}


\subsection{Unbiased LTR for Continuous Labels} \label{cont_ltr}


With continuous user feedback,  the aforementioned unbiased LTR method cannot work. To circumvent the limitation of the strong assumption between unbiased relevance labels and biased user action labels, we propose a novel unbiased LTR method from the perspective of pairwise learning, which can work for both categorical and continuous labels.

In the pairwise setting, the loss function is defined on item pairs. Let $\hat{y}_i=f(u,x_i)$ denotes the  prediction score for user item pair $(u,x_i)$, $L(\hat{y}_i,r_i, \hat{y}_j ,r_j)$ denotes the pairwise loss function and $P(r_i \!\!> \!\! r_j,u,x_i,x_j)$ denotes the probability distribution that user $u$ prefers item $x_i$ over item $x_j$, the risk function and the the ranker learned through minimizing the empirical risk function are defined in Equation \ref{eq:6} and Equation \ref{eq:7} respectively. Similar to unbiased LTR for categorical labels, we also consider positive relevance pairs only as negative and neutral pairs should not be involved to optimize common ranking metrics such as DCG and ARP.

\vspace{-0.4cm}
\begin{equation} \label{eq:6}
 R_{rel}(f)  =    \int_{}{} L(\hat{y}_i,r_i, \hat{y}_j ,r_j) dP( r_i  >  r_j,  u,  x_i,  x_j ) 
\end{equation}
\vspace{-0.8cm}

\begin{equation} \label{eq:7}
\hat{f}_{rel}  =   \arg\min_{f}   \sum_{u, x_i, x_j,  r_i >  r_j}  L(\hat{y}_i,r_i, \hat{y}_j ,r_j)
\end{equation}

\subsubsection{Pairwise PBM based Unbiased LTR} 
To model the relation between unbiased relevance pair  $(r_i,r_j)$ and biased continuous label pair $(c_i,c_j)$, we extend the PBM to the pairwise setting. Specifically, we assume that the examination of item $x_i$ and $x_j$ are independent and if  $x_i$ and $x_j$ are both examined, then $x_i$ exhibits a larger continuous label than $x_j$ if and only if $x_i$ is more relevant than $x_j$, which is:
\begin{equation}
P(c_i \! > \! c_j,e_i=1,e_j=1 | u,x_i,x_j,i,j) \! = \! \theta_i \theta_j \gamma_{u,x_i,x_j}
\end{equation}
, where $\gamma_{u,x_i,x_j}=P(r_i>r_j|u,x_i,x_j)$. Then we can derive the risk function based on the positive continuous label pairs: 
\begin{align}
&\!\!\! \!  R_{unbiased}(f) \nonumber  \\
&\!\!\! \! \!\!= \!\! \int_{}{}  \frac {L(\hat{y}_i, \! c_i, \hat{y}_j, \! c_j)} {\theta_i \theta_j } \! P(  e_j  \! =  \! 1|  u,\! i, \! j, \! x_i, \! x_j ,  c_i > c_j ) dP( c_i \!\! > \!\! c_j, \! u, \! x_i, \! x_j, \! i , \!j ) \! \nonumber  \\
& \!\!\! \! \!\!= \!\! \int_{}{}  \frac {L(\hat{y}_i, \! c_i, \! \hat{y}_i, \! c_j)} {\theta_i \theta_j }  dP( c_i \!\! > \!\! c_j, \! u, \! x_i, \! x_j,i,j ,e_i=1,e_j=1\!) \! \nonumber  \\
&\!\!\! \! \!\!= \!\! \int_{}{}  \frac {L(\hat{y}_i, \! c_i, \! \hat{y}_i, \! c_j)} {\theta_i \theta_j }  dP( r_i \!\! > \!\! r_j, \! u, \! x_i, \! x_j,i,j ,e_i=1,e_j=1\!) \!  \nonumber  \\
&\!\!\! \! \!\!= \!\! \int_{}{}  \frac {L(\hat{y}_i, \! r_i, \! \hat{y}_j, \! r_j)} {\theta_i \theta_j }  dP( r_i \!\! > \!\! r_j, \! u, \! x_i, \! x_j,i,j) P(e_i \! = \! 1|i)P(e_j \!= \! 1|j) \! \nonumber  \\
&\!\!\! \! \!\!= \!\! \int_{}{}  {L(\hat{y}_i, \! r_i, \! \hat{y}_i, \! r_j)} dP( r_i \!\! > \!\! r_j, \! u, \! x_i, \! x_j \!) \! = R_{rel}(f)
\end{align}

Based on the justification that $R_{unbiased}$ is an unbiased estimator of $R_{rel}$, we can learn an unbiased ranker through minimizing the empirical risk function ${loss_{IPW}}$ shown in Equation 10, where $L_{ij}=L(\hat{y}_i, \! c_i, \! \hat{y}_j , \! c_j)$. It is worth noting that samples with positive labels and zero labels are separately handled explicitly for better precision in the empirical risk function, as the posterior probability of examination is different for samples with different labels. In detail, the posterior examination probability is definitely 1 for items with a positive label but is uncertain for items with zero labels.
\begin{align}
& \!\!\!\!\!loss_{IPW} = \!\!\!\!\!\!\!\!\!\!\!\!\!\!\!\!\!\!   \sum_{u, x_i, x_j,  i,   j,  c_i >  c_j,c_j>0}  \! \frac {L_{ij}} {\theta_i \theta_j} \! +  \!\!\!\!\!\!   \sum_{u, x_i, x_j,  i,   j, c_i > c_j,c_j=0}  \!\!\!\!  \frac {L_{ij}h_{ij}} {\theta_i \theta_j} \!,  \!\!
\end{align}
where $h_{ij}=P(e_j=1|u,x_i,x_j,i,j,c_i>c_j,c_j=0)$.



\begin{figure*}[!htbp]
\setlength{\abovecaptionskip}{-0.1cm} 
\setlength{\belowcaptionskip}{-0.2cm}
	\centering
	\includegraphics[width=0.9\textwidth, trim=40 490 40 70,clip]{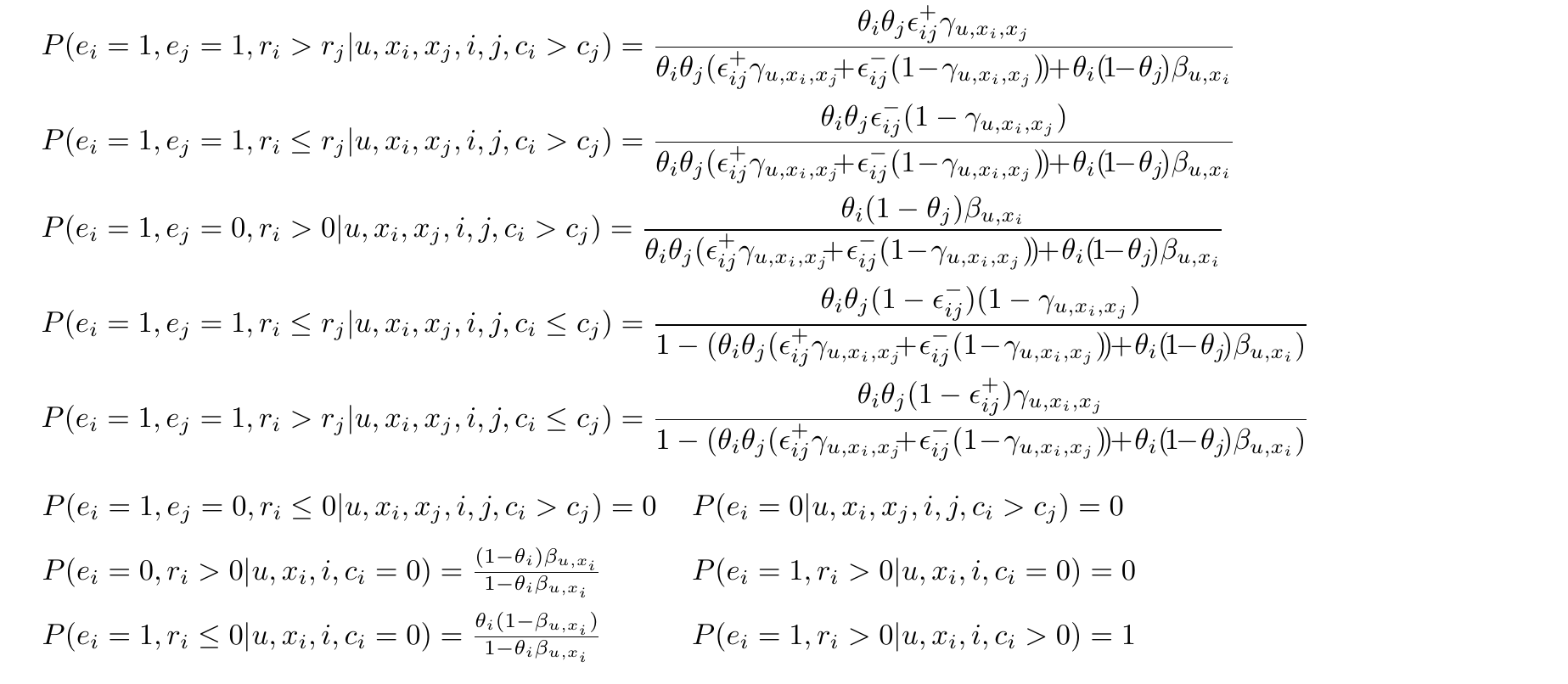}
	\caption{Formulas in Expectation Step}
	\label{estep}
\end{figure*}


\subsubsection{Pairwise Trust Bias based Unbiased LTR.} The pairwise PBM based unbiased LTR method assumes that if $x_i$ is more relevant than $x_j$ and they are both examined, $x_i$ will exhibit a larger continuous label than $x_j$. However, according to the user study in \cite{joachims2017accurately}, users tend to trust higher-ranked items more, so higher-ranked items exhibit larger click ratios even if items in different positions are all examined and equally relevant, which is called the trust bias. Moreover, the continuous feedback value usually follows a distribution rather than a specific mapping from the relevance label, which engenders even more complex relations between biased label pairs and unbiased relevance pairs. To estimate the relation between unbiased relevance pair $(r_i,r_j)$ and biased continuous label pair $(c_i,c_j)$ more accurately, we incorporate the trust bias explained above and extend it to the pairwise setting in this paper. Since the trust bias only depends on the exposed position, we formulate the pairwise trust bias in Equation 11 and Equation 12 and assume that $P(c_i \! > \! c_j|e_i=1,e_j=1, r_i>r_j,u,x_i,x_j,i,j)$ and $P(c_i \! > \! c_j|e_i=1,e_j=1, r_i \leq r_j, u,x_i,x_j,i,j)$ are independent of $(u,x_i,x_j)$. Specifically, for a positive relevance pair $r_i>r_j$, the probability of exposing a positive continuous label pair $c_i >c_j$ is not 1 but $\epsilon_{ij}^+$, while a non-positive relevance pair also has the probability $\epsilon_{ij}^-$ to exhibit a positive continuous label pair. $\epsilon_{ij}^+$ and $\epsilon_{ij}^-$ are only position-dependent and follows the constraint $0<\epsilon_{ij}^-<\epsilon_{ij}^+<1$.
\vspace{-0.1cm}
\begin{align}
P(c_i \! > \! c_j|e_i=1,e_j=1, r_i>r_j,i,j) \! = \epsilon_{ij}^+ \\ 
P(c_i \! > \! c_j|e_i=1,e_j=1, r_i \leq r_j,i,j) \! = \epsilon_{ij}^-
\end{align}


Then we can calculate the probability of a positive continuous label pair. As shown in Equation 13, there are two cases to expose a positive continuous label pair $c_i \!\! > \!\! c_j$. First, if $x_j$ is examined, the relation between $c_i$ and $c_j$ depends on the relation between $r_i$ and $r_j$ and the trust bias. Second, if $x_j$ is not examined, the positive pair $c_i \! > \! c_j$ reduces to the positive data point $c_i \! > \! 0$, which only depends on the examination and relevance of $x_i$ according to PBM.
\begin{align}
\!\!\!\! & P(c_i  >  c_j|u,x_i,x_j,i,j)   \nonumber  \\
&  =  P(c_i > c_j, e_i=1, e_j=1|u,x_i,x_j,i,j)  \nonumber  \\
& + P(c_i>0, e_i=1, e_j=0|u,x_i,x_j,i,j)  \nonumber \\
 & = \!  \theta_i \theta_j  ( \epsilon_{ij}^+ \gamma_{u,x_i,x_j}  +  \epsilon_{ij}^-(1  -  \gamma_{u,x_i,x_j}))  +  \theta_i (1  -  \theta_j) \beta_{u,x_i}
\end{align}
, where $\beta_{u,x_i}$ \! = \! $P(r_i>0|u,x_i)$ is the relevant probability of $(u,x_i)$.

As we can no longer deduce the state of $r_i \! > \! r_j$ from the condition of $c_i \! > \! c_j,e_i \! = \! 1,e_j \!= \!1$, $R_{rel}$ could not be derived from $R_{unbiased}$ with the introduction of pairwise trust bias. To tackle this issue, we calculate the probability of positive relevance pair given the positive continuous label pair and true examination, which is:
\begin{align} \label{eq:14}
\!\!\!\!\! & P(r_i \! > \! r_j|u,x_i,x_j,i,j,c_i>c_j,e_i=1,e_j=1)   \nonumber  \\
&  =\! \frac {P(c_i\!\! >\!\! c_j| e_i \! = \! 1,\! e_j \! = \! 1,\!r_i \!\!>\! \! r_j, \! u, \! x_i,\!x_j,\! i, \!j)P(r_i \!\! > \!\! r_j|u,\! x_i,\! x_j)} {P(c_i\! >\! c_j,\! e_i=1,\! e_j=1|u,x_i,x_j,i,j)}   \nonumber  \\
& = \! \frac {\epsilon_{ij}^+ \gamma_{u,x_i,x_j}} {\epsilon_{ij}^+ \gamma_{u,x_i,x_j}  +  \epsilon_{ij}^-(1  -  \gamma_{u,x_i,x_j})} = m_{ij}
\end{align}

Finally, we define a Bayes-IPW loss based on $loss_{IPW}$ and $m_{ij}$, which compensates and corrects for both the position-based examination bias and trust bias. One can see from Equation 15 and Equation 14 that the Bayes-IPW loss reduces to the $loss_{IPW}$ in the noisy free case of pairwise PBM, i.e., when $\epsilon_{ij}^+=1$ and $\epsilon_{ij}^-=0$.
\begin{align}
\!\!\!\!\!\!\!  loss_{Bayes-IPW} \! = \!  \arg\min_{f} \!\!\!\!\!\!\!   \sum_{u, x_i, x_j,  i,   j, \atop  c_i >  c_j,c_j>0} \!\!\!\!\!\!\!  \frac {L_{ij}m_{ij}} {\theta_i \theta_j} \! 
 +  \!\!\!\!\!\!\!\!\!  \sum_{u, x_i, x_j,  i,   j, \atop c_i > c_j,c_j=0}  \!\!\!\!\!\!\!\!\!  \frac {L_{ij}m_{ij}h_{ij}} {\theta_i \theta_j} \!,  \!\!
\end{align}
where ${h_{ij}}$ is calculated as follows:
\begin{align}
\!\!\!\!\!\!\!\!\!\! & h_{ij} = P(e_j=1|u,x_i,x_j,i,j,c_i>c_j,c_j=0)  \nonumber  \\
& = \!\! \frac {P(e_j \!= \! 1|u, \! x_j,\!j,\! c_j \!\! = \! 0) P(c_i \!\! > \!\! c_j|u,\!x_i,\!x_j,\!i,\!j,\!e_j \! = \! 1,\! c_j \! = \! 0)} {P(c_i \! > \! c_j|u,x_i,x_j,i,j,c_j \! = \! 0)}  \nonumber  \\
& = \!\! \frac {\theta_i \theta_j^-  ( \epsilon_{ij}^ + \gamma_{u,x_i,x_j} \!\!\! + \! \epsilon_{ij}^-(1 \! - \!\! \gamma_{u,x_i,x_j})\!)}
 {\!  \theta_i \theta_j^- \!\!\ ( \epsilon_{ij}^+ \gamma_{u,x_i,x_j} \!\!\! + \! \epsilon_{ij}^-(1 \!\! - \! \gamma_{u,x_i,x_j})\!) \!\! + \!\! \theta_i (1 \!\! - \! \theta_j^- \! ) \beta_{u,x_i}}\!,
\end{align}
where $\theta_j^-$ denotes the posterior examination probability when the continuous label of item $x_j$ is known to be 0.

\subsubsection{Direct Optimization for Ranking Metrics.}As pairwise loss \cite{burges2005learning} focuses on minimizing the number of pairwise errors and neglecting the relative importance of item pairs with different relevance gaps, which does not match well with common performance metrics such as DCG and ARP. To directly optimize the ranking metrics in the model, we refine the loss function following the practice of LambdaRank \cite{burges2006learning}. The main idea of LambdaRank is to incorporate a delta NDCG to directly optimize the evaluation metric of NDCG, where delta NDCG denotes the difference between NDCG scores if item  $x_i$ and $x_j$ are swapped in the ranking list \cite{burges2006learning,burges2010ranknet}.
As the unbiased true relevance is not available, we assume that the biased feedback is positively correlated to the unbiased relevance, and refine the pairwise loss $loss_{Bayes-IPW}$ to   $loss_{opt}$ as follows:
\begin{align}
  \!\!\! & loss_{opt} = \!  \!\!\!\!\!\!\!   \sum_{u, x_i, x_j,  i,   j, \atop c_i >  c_j,c_j>0}  \!\!\!\!\!\!\!\!\!  \frac {L_{ij} m_{ij} | \Delta Z_{ij}| } {\theta_i \theta_j} \!  
  +  \!\!\!\!\!\!\!\!\!\!   \sum_{u, x_i, x_j,  i,   j, \atop  c_i > c_j,c_j=0}  \!\!\!\!\!\!\!  \frac {L_{ij} m_{ij}h_{ij} | \Delta Z_{ij}| } {\theta_i \theta_j} \!,  \!\!
\end{align}
where $\Delta Z_{ij}$ is the difference between evaluation metrics based on the biased label if item $x_i$ and $x_j$ are swapped in the ranking list. 

Thus, if we can estimate the parameters of $\theta_i$, $\theta_i^-$, $\epsilon_{ij}^+$ and $\epsilon_{ij}^-$, we can learn an unbiased ranker $\hat{f}_{unbiased}$ through minimizing the loss function. In this paper, we employ a regression-based Expectation-Maximization (EM) method \cite{wang2018position} to estimate the parameters. The estimation process would be described in the next section.

\subsection{Supervision Label Synthesis for Multi-Objective Applications} \label{subsection:3}

Some applications, such as content recommender systems, often consider multiple live metrics simultaneously \cite{tang2020progressive}. For example, we would like to maximize the user’s video view count, dwell time, and other satisfaction-related metrics (e.g., liking and sharing) in the video recommendation scenario. To model multiple actions and maximize multiple live metrics simultaneously, conventional recommender systems often train multiple rankers to predict the absolute value of multiple objectives and combine these predictions for final ranking. However, the optimization target in such a training paradigm differs a lot from the ranking metrics valuing at the relative order of top-ranked items rather the absolute scores. Moreover, if we train multiple rankers through different unbiased LTR models, we could not combine predictions in a theoretically sound way as predictions of different unbiased LTR models might not be in the same scale. 

To address these issues and achieve better multi-objective optimization performance, we propose to define a synthesized label based on diverse user actions so as to directly model multiple live metrics in one single model. As shown in Equation \ref{eq:18}, the synthesized label merges multiple user actions (e.g., click, dwell time, and satisfaction) through a combination function $\Phi$, where $M$ is the number of user actions considered in live metrics. The combination function is usually set based on business goals, and its form can be fairly flexible to capture complex relations between live metrics. For instance, weighted summation and weighted multiplication are commonly used in industrial recommender systems. With the synthesized continuous label, we can naturally apply the method described in section \ref{cont_ltr}.

\begin{equation} \label{eq:18}
c = \Phi(action_1, action_2, ... , action_M)
\end{equation}

There are some advantages to combining multiple user behaviors to the synthesized label. First, the gap between the training objective and the live metrics can be bridged through learning from the synthesized label, as the synthesized label is closely related to the ultimate business goals. Second, we can obtain the unbiased prediction of the live metric from the single model and employ it for final ranking directly without combination, which is more convenient and efficient.

\section{Estimation via Expectation-Maximization}

To estimate parameters of $\theta_i$, $\theta_i^-$, $\epsilon_{ij}^+$ and $\epsilon_{ij}^-$ introduced in the last section and learn the unbiased ranker, we extend the regression-based EM method \cite{wang2018position} to the pairwise setting in this paper. In the procedure of regression-based EM algorithm, parameters are estimated by iterating over the Expectation steps and Maximization steps until convergence. In this section, we illustrate the process of the Expectation step and the Maximization step respectively.

\begin{figure*}[!htbp]
\setlength{\belowcaptionskip}{-0.3cm}
\setlength{\abovecaptionskip}{-0.05cm} 
	\centering
	\includegraphics[width=0.94\textwidth, trim=50 533 0 70,clip]{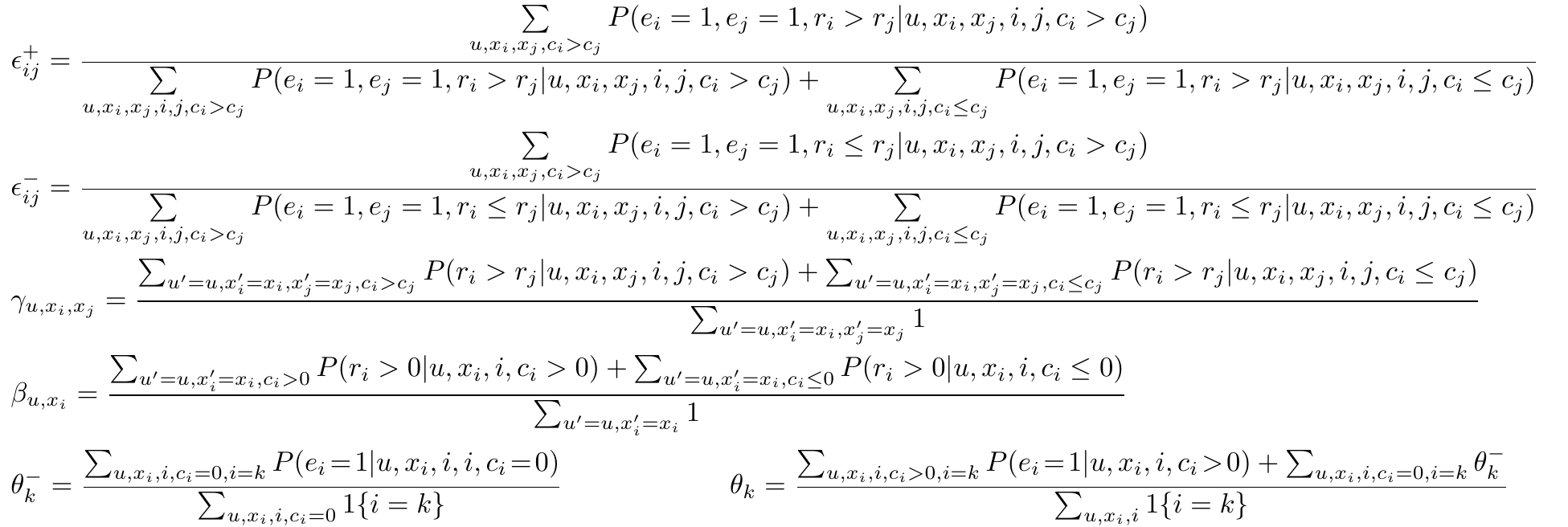}
	\caption{Formulas in Maximization Step}
	\label{mstep}
\end{figure*}
\vspace{-0.1cm}

\subsection{Expectation Step}
In the Expectation step, we need to estimate the distribution of hidden variable $e_i$, $e_j$ and relevance pair $(r_i,r_j)$ given parameters $\theta_i$, $\epsilon_{ij}^+$ and $\epsilon_{ij}^-$ and $\gamma_{u,x_i,x_j}$. To achieve the goal, we first calculate the joint probability of all hidden variables shown in \autoref{estep} , where all formulas  are derived directly from Bayes rules. For instance, in the first equation, we have:
\begin{align}
& P(e_i=1,e_j=1,r_i>r_j|u,x_i,x_j,i,j,c_i>c_j)  \nonumber  \\
& = \frac {P(c_i \! > \! c_j|e_i \! = \! 1,e_j \! = \! 1,r_i \! > \!  r_j) P(e_i \! = \! 1,e_j \! = \! 1,r_i \! > \! r_j)} {P(c_i \! > \! c_j|u,x_i,x_j,i,j)}  \nonumber  \\
& = \frac { \theta_i \theta_j \epsilon_{ij}^+ \gamma_{u,x_i,x_j} } { \theta_i \theta_j  ( \epsilon_{ij}^+ \gamma_{u,x_i,x_j} \!\!\! + \! \epsilon_{ij}^-(1 \! - \! \gamma_{u,x_i,x_j})\!) \!\! + \! \theta_i (\! 1 \!\! - \! \theta_j\!) \beta_{u,x_i}}
\end{align}

It is worth noting that we estimate hidden variables $e_i$, $e_j$ and true relevance $r_i>0$ in the pointwise fashion, which is shown in the last 4 equations in \autoref{estep}, as they only depend on a single item. After calculating the joint probability, we can then calculate the marginal probability of hidden variables, which will be used in the Maximization step. For instance, the marginal probability $P(e_i=1|c_i=0)$  equals to the sum of $P(e_i=1,r_i>0 | c_i=0)$ and $P(e_i=1,r_i \leq 0 | c_i=0)$.

\subsection{Maximization Step}
In the maximization step, all parameters are updated to minimize the loss function, given the training samples and the posterior probabilities from the Expectation step. All formulas for parameter updating are presented in \autoref{mstep}.
In the procedure of standard EM algorithm, $\gamma_{u,x_i,x_j}$ and $\beta_{u,x_i}$  are updated following the $3rd$ equation and $4th$  equation respectively, which requires that  $(u,x_i,x_j)$ should repeat and be shown in different position pairs. As samples of user-item-pair $(u,x_i,x_j)$ are highly sparse and synthesized label pairs $(c_i,c_j)$ are noisy in large-scale real-world systems, it is extremely difficult to minimize loss through free parameters $\gamma_{u,x_i,x_j}$ and $\beta_{u,x_i}$. To address the problem, we apply the regression-based EM \cite{wang2018position} algorithm to estimate parameters of $\gamma_{u,x_i,x_j}$ and $\beta_{u,x_i}$ via learning a regression function. 
Specifically, we assume that there are feature vector $X_{u,x_i,x_j}$  and $X_{u,x_i}$  representing the sample of user-item-pair $(u,x_i,x_j)$ and user-item $(u,x_i)$ respectively, we use function $g$ to compute the relevance preference $\gamma_{u,x_i,x_j} \!\!= \! g(X_{u,x_i,x_j })$ and function $h$ to compute absolute relevance $\beta_{u,x_i } \! = \! h(X_{u,x_i })$. Thus, we aim to find regression functions $g(X)$ and  $h(X)$ in the maximization step to minimize the loss function, given training samples and the distributions of hidden variables calculated in the Expectation step. For instance, we can regress the feature vector  $X_{u,x_i,x_j}$ to the probability $P( r_i \!\! > \!\! r_j|u,x_i,x_j,i,j,c_i,c_j )$. Similar to \cite{wang2018position}, we convert such a regression problem to a classification problem based on Bernoulli sampling. Specifically, we sample a binary relevance pair label $\gamma \in \{0,1\}$ and a binary relevance label $\beta \! \in \! \{0,1\}$ according to $P( r_i \!> \! r_j|u,x_i,x_j,i,j,c_i,c_j )$ and $P( r_i \!>0|u,x_i,i,c_i )$ respectively. Then we can adopt classification models to learn $g(X)$ and $h(X)$ based on training set $\{X,\gamma\}$ and $\{X,\beta\}$. It is flexible to use any classification model for $g(X)$ and $h(X)$ and we choose the commonly used DNN model in this paper.

	
Note that the original EM algorithm updates parameters in the Maximization step through calculation on the whole data set, which imposes challenges for industrial systems with large-scale training data. To address the problem, we employ the mini-batch EM following the idea of online EM \cite{cappe2009line}. Accordingly, we calculate the formulas in Figure 2 based on data in one single mini-batch, and update parameters incrementally in each batch as follows:
\begin{equation}
\epsilon_{ij}^+ = \epsilon_{ij}^+ \times (1-\alpha) + \widehat {\epsilon_{ij}^+} \times \alpha
\end{equation}
where $\alpha$ is the scheduled learning rate of $\epsilon_{ij}^+$, $\widehat {\epsilon_{ij}^+} $ is the estimation of $\epsilon_{ij}^+$ in the current batch.

\begin{table*}
\setlength\tabcolsep{3pt}
\begin{tabular}{lccc|ccc}
\hline
& \multicolumn{3}{c}{Yahoo Dataset } &  \multicolumn{3}{c}{MovieLens} \\
& Reward@3 & Reward@5 & Reward@10 & Reward@3 & Reward@5 & Reward@10 \\
\hline
Our Method ($loss_{opt}$) & \textbf{ 2.3319} $\pm$ 6.9e-3 & \textbf{ 2.7185} $\pm$ 7.5e-3 & \textbf{ 2.9735} $\pm$ 5.9e-3 & 5.8553 $\pm$ 1.1e-2 & \textbf{ 7.6690} $\pm$ 1.5e-2 &  \textbf{9.1400} $\pm$ 1.6e-2 \\
Our Method ($loss_{Bayes-IPW}$)  & 2.3046 $\pm$ 8.5e-3 & 2.6864 $\pm$ 7.5e-3 & 2.9423 $\pm$ 8.0e-3 & 5.8415 $\pm$ 1.4e-2 & 7.6486 $\pm$ 7.8e-3 & 9.1217 $\pm$ 9.7e-3  \\ 
Our Method ($loss_{IPW}$) & 2.2447 $\pm$ 9.9e-3 & 2.6239 $\pm$ 1.0e-2 & 2.8828 $\pm$ 8.9e-3  & \textbf{ 5.8584} $\pm$ 2.0e-2 & 7.6635 $\pm$ 2.1e-2 & 9.1285 $\pm$ 2.5e-2 \\

\hline
\hline
Sum-Synthesize & 2.2599 $\pm$ 5.0e-3 & 2.6380 $\pm$ 8.1e-3 & 2.8913 $\pm$ 8.7e-3 & 5.7863 $\pm$ 2.5e-2 & 7.5704 $\pm$ 3.1e-2 & 9.0289 $\pm$ 3.5e-2 \\ 
PAL-Synthesize & 2.2612 $\pm$ 8.7e-3 & 2.6388 $\pm$ 6.0e-3 & 2.8914 $\pm$ 5.6e-3 & 5.7484 $\pm$ 3.1e-2 & 7.5271 $\pm$ 3.7e-2 & 8.9682 $\pm$ 4.0e-2 \\
Sum-Click & 2.2776 $\pm$ 1.2e-2 & 2.6590 $\pm$ 9.7e-3 & 2.9146 $\pm$ 8.9e-3 & 5.7283 $\pm$ 2.1e-2 & 7.5570 $\pm$ 2.3e-2 & 9.0264 $\pm$ 2.8e-2 \\
PAL-Click & 2.2907 $\pm$ 9.9e-3 & 2.6719 $\pm$ 6.6e-3 & 2.9274 $\pm$ 6.7e-3  & 5.7385 $\pm$ 1.9e-2 & 7.5838 $\pm$ 1.8e-2 & 9.0586 $\pm$ 1.9e-2 \\


PAL-Click + PAL-DwellTime & 2.2707 $\pm$ 7.7e-3 & 2.6504 $\pm$ 7.3e-3 & 2.9058 $\pm$ 6.9e-3 & 5.7569 $\pm$ 2.7e-2 & 7.5491 $\pm$ 3.5e-2 & 9.0031 $\pm$ 3.7e-2 \\
Lower Bound (pointwise DNN) & 2.2548 $\pm$ 4.4e-3 & 2.6353 $\pm$ 6.5e-3 & 2.8892 $\pm$ 5.3e-3 & 5.7106 $\pm$ 1.3e-2 & 7.5392 $\pm$ 1.6e-2 & 9.0113 $\pm$ 1.9e-2  \\
Lower Bound (LambdaRank) & 2.1619 $\pm$ 1.2e-2 & 2.5370 $\pm$ 8.5e-3 & 2.7957 $\pm$ 8.4e-3 & 5.6800 $\pm$ 2.1e-2 & 7.4805 $\pm$ 2.9e-2 & 8.9387 $\pm$ 4.7e-2 \\

\hline
\end{tabular}
\caption{Experiment Results on Continuous Labels}
\label{table1}
\end{table*}

\section{Experiments} \label{section:5}

In this section, we first describe the experiment data and experiment setup, then conduct extensive offline and online experiments to answer the following research questions:

\begin{itemize}
\item RQ1: How does the proposed method perform compared with the SOTA debiasing methods on continuous labels?
\item RQ2: How does the proposed method perform compared with the SOTA unbiased LTR methods on categorical labels?
\item RQ3: Whether the proposed method is robust in different severity of position bias and different relative weight of continuous labels over categorical labels?
 \item RQ4: How does the proposed method perform in real-world recommender systems?
\end{itemize}

\subsection{Experiment Data}
To answer the aforementioned research questions and evaluate the effectiveness of the proposed method in both recommendation and searching scenarios, we conduct experiments on both LTR and recommendation public benchmark datasets, which are:
\begin{itemize}
    \item LTR dataset: Yahoo! LETOR set 1\footnote{https://webscope.sandbox.yahoo.com/catalog.php?datatype=c} and MSLR-Web30K \footnote{https://www.microsoft.com/en-us/research/project/mslr/} are two of the largest public LTR benchmark datasets, which contain multiple query document pairs represented by feature vectors and 5-level relevance labels. Specifically, there are 29.9k queries with 710k documents in Yahoo Dataset and 31.2k queries in MSLR-Web30K. We adopt the training, validation, testing splits in these corpora. 
    \item Recommendation dataset: MovieLens \footnote{https://files.grouplens.org/datasets/movielens/ml-1m-README.txt} is a recommendation benchmark dataset containing 1M unbiased ratings of 6k users on 3.9k movies. As each user and movie is only represented by an id without other information in MovieLens, we generate feature vectors for all user movie pairs in the following steps. First, we split samples of each user into training(50\%), validation(25\%), and testing(25\%) set based on the time order. Then we train a BPR \cite{rendle2012bpr} model based on the training set to obtain a 60-dimensional feature vector for each user and each item. Finally, we concatenate the feature vector of user, item, and dot of user and item to generate a 180-dimensional feature vector for each user-item pair.
\end{itemize}

As there is no user action in these datasets, we consider each query or user as a user request and generate the click and dwell-time labels to simulate biased feedback.

\subsubsection{Click Simulation}
 We follow the procedures in \cite{ai2018unbiased} to sample clicks. First, we train a Ranking-SVM \footnote{http://www.cs.cornell.edu/people/tj/svm\_light/svm\_rank.html} model  using 1\% of the training data with relevance labels to generate the initial ranking list for each user. Then we sample click label $click_{u,x_i,i}$ uniformly by simulating the browsing process of PBM as follows:
 \begin{align} \label{eq:21}
&     click_{u,x_i,i}   \sim  U(P(c_i=1|u,x_i,i)), \nonumber \\
&     P(c_i=1|u,x_i,i)=P(e_i=1|i)P(r_i=1|u,x_i), 
 \end{align}
where $P(e_i=1|i)=\theta_i^\eta$ denotes the probability of examination, $\theta_i$ is the position bias of click at position $i$ determined by the initial Ranking-SVM ranker and $\eta$ controls the severity of position bias. $\theta_i$ is obtained from an eye-tracking experiment in \cite{joachims2017accurately} and $\eta$ is set as 1 in default.
The probability of an item $x_i$ to be perceived as relevant is calculated as:

\begin{center}
$P(r_i=1|u,x_i)= \epsilon + (1-\epsilon) \frac {2^{y_{u,x_i}}-1} {2^{y_{max}}-1},$
\end{center}

\noindent where $y_{u,x_i} \!\! \in \!\! [0,4]$ is the 5-level relevance label and $y_{max}$ is the highest level of 4. We map ratings in MovieLens from [1-5] to [0-4] as relevance labels. $\epsilon$  models click noise so that irrelevant items $(y_{u,x_i} \!\! = \! 0)$ have non-zero click probability. $\epsilon$ is set to 0.1 by default.

\begin{table*}
\setlength\tabcolsep{4pt}
\begin{tabular}{lccc|ccc}
\hline
& \multicolumn{3}{c}{Yahoo Dataset } &  \multicolumn{3}{c}{MSLR-Web30K} \\
& NDCG@3 & NDCG@5 & NDCG@10 & NDCG@3 & NDCG@5 & NDCG@10 \\
\hline
Upper Bound &0.6960 $\pm$ 1.4e-3 &0.7175 $\pm$ 1.3e-3 &0.7644 $\pm$ 1.0e-3 &0.3960 $\pm$ 2.3e-3 &0.4047 $\pm$ 1.8e-3 &0.4286 $\pm$ 1.8e-3 \\
\hline
\hline
Our Method ($loss_{opt}$) &0.6895 $\pm$ 2.5e-3 & \textbf{0.7115} $\pm$ 2.1e-3 & \textbf{0.7589} $\pm$ 1.4e-3 & \textbf{0.3811} $\pm$ 2.6e-3 & \textbf{0.3899} $\pm$ 1.7e-3 & \textbf{0.4132} $\pm$ 1.0e-3 \\
Our Method ($loss_{Bayes-IPW}$)  & \textbf{0.6900} $\pm$ 2.4e-3 & \textbf{0.7116} $\pm$ 1.7e-3 & \textbf{0.7589} $\pm$ 1.5e-3  &0.3800 $\pm$ 2.5e-3 &0.3888 $\pm$ 1.8e-3 &0.4125 $\pm$ 1.1e-3 \\ 
Our Method ($loss_{IPW}$) &0.6860 $\pm$ 1.6e-3 &0.7079 $\pm$ 1.3e-3 &0.7557 $\pm$ 1.2e-3 &0.3784 $\pm$ 1.4e-3 &0.3882 $\pm$ 1.4e-3 &0.4119 $\pm$ 1.5e-3 \\

\hline
\hline

DLA \cite{ai2018unbiased} &0.6892 $\pm$ 2.1e-3 &0.7106 $\pm$ 2.2e-3 &0.7578 $\pm$ 1.6e-3 &0.3792 $\pm$ 2.7e-3 &0.3870 $\pm$ 2.4e-3 &0.4102 $\pm$ 2.9e-3 \\
Regression EM \cite{wang2018position}   &0.6893 $\pm$ 2.1e-3 &0.7108 $\pm$ 2.1e-3 &0.7577 $\pm$ 1.6e-3 &0.3707 $\pm$ 3.9e-3 &0.3766 $\pm$ 3.7e-3 &0.3977 $\pm$ 3.6e-3 \\
Pairwise Debiasing \cite{hu2019unbiased}  &0.6600 $\pm$ 3.5e-3 &0.6847 $\pm$ 2.8e-3 &0.7365 $\pm$ 2.0e-3 &0.3481 $\pm$ 3.3e-3 &0.3623 $\pm$ 2.8e-3 &0.3912 $\pm$ 2.4e-3 \\
Lower Bound (LambdaRank)  &0.6541 $\pm$ 2.5e-3 &0.6790 $\pm$ 2.2e-3 &0.7320 $\pm$ 1.7e-3 &0.3433 $\pm$ 2.2e-3 &0.3579 $\pm$ 2.1e-3 &0.3874 $\pm$ 1.7e-3 \\
\hline
\end{tabular}
\caption{Experiment Results on Categorical Labels}
\label{table2}
\end{table*}

\subsubsection{Dwell-Time Simulation}
Then we sample continuous dwell-time labels based on sampled click data as only clicked samples have non-zero dwell-time in practice. As shown in \cite{yin2013silence}, the dwell-time is correlated with the relevance and item length, and dwell-time of items with the same length satisfies a log-Gaussian distribution. So we analyze dwell-time of videos around 100s in a real-world video recommender system and also find a log-Gaussian distribution. Furthermore, we assume that dwell-time in different relevance levels follows different log-Gaussian distribution and the overall log-Gaussian of dwell-time is composed of these distributions. Then we train a GMM(Gaussian Mixture Model) \cite{reynolds2009gaussian} on log scaled dwell-time to obtain 5 Gaussian distributions for different relevance level and sample dwell-time as follows:



\begin{equation} \label{eq:22}
dt_{u,x_i,i} = click_{u,x_i,i} \cdot e^{\omega_{u,x_i}} ,  \quad  \omega_{u,x_i} \! \sim \! N(\mu_{y_{u,x_i}}, \sigma_{y_{u,x_i}}),  
\end{equation}
where $click_{u,x_i,i}$ is the sampled click label, $\omega_{u,x_i}$ denotes the sampled log-scaled dwell-time after click,  $\mu_{y_{u,x_i}}$ and $\sigma_{y_{u,x_i}}$ are mean and standard deviation in different relevance level obtained by GMM.

\subsection{Experiment Setup}
In this paper, we perform experiments on the synthesized label in Subsection \ref{subsection53} and the click label in Subsection \ref{subsection54} to evaluate the performance of our method on continuous and categorical labels respectively. Based on Subsection \ref{subsection:3}, we adopt the commonly used weighted summation function in typical recommendation scenarios and define the synthesized label $c_{u,x_i,i}$ as follows:

\begin{equation}
   c_{u,x_i,i} = click_{u,x_i,i} + \frac{dt_{u,x_i,i}}{e^\delta},  \nonumber
\end{equation}
where $\delta$ controls the weight of continuous label in the synthesized label. $\delta$ is set to 3 by default to keep the similar weight of dwell-time as our real-world recommender system.

In the experiment, we adopt a DNN model with [512, 256, 128] hidden units, ELU activation, Adagrad optimizer, cross-entropy loss for binary classification ranker and pairwise ranker, and MSE loss for regression ranker. Besides, we tune the learning rate in [5e-3, 5e-2] carefully for each method. For a fair comparison, we train and evaluate each method 10 times and report the mean and standard deviation of evaluation metrics. To evaluate the unbiased LTR performance of our method, we adopt NDCG@k as the evaluation metric for experiments on categorical labels. As for continuous labels, we adopt the commonly used weighted summation function in typical recommendation applications to combine click and dwell-time, and define the evaluation metric $Reward@k$ as follows to simulate the combined live metric:

\begin{equation}
  Reward@k=\sum_{i=1}^{k}{click^{\prime}_{u,x_i,i} + \frac{dt^{\prime}_{u,x_i,i}}{e^\delta}},  \nonumber
\end{equation}
where $click^{\prime}_{u,x_i,i}$ and $dt^{\prime}_{u,x_i,i}$ are estimated click and dwell-time feedback at ranked position $i$. Unlike $click_{u,x_i,i}$, the position $i$ here is determined by the final unbiased LTR ranker rather than the initial Ranking-SVM ranker. Specifically, $click^{\prime}_{u,x_i,i}$ and $dt^{\prime}_{u,x_i,i}$ are sampled based on Equation \ref{eq:21} and Equation \ref{eq:22} respectively.

\subsection{Experiments on Continuous Labels (RQ1)} \label{subsection53}
We first conduct experiments on the continuous synthesized label of Yahoo dataset and MovieLens to evaluate the performance of our method in real-world applications. For a comprehensive comparison, we train a pointwise DNN and LambdaRank model with the synthesized label without debiasing as the lower bound.  Moreover, we remove the pairwise trust bias correction and loss refinement for direct metrics optimization from our method to explore their effectiveness. Note that the SOTA industrial recommender systems train multiple pointwise models to predict multiple objectives respectively and employ PAL \cite{guo2019pal} or sum-based shallow tower \cite{zhao2019recommending} for debiasing. We consider the following baseline methods:

\begin{itemize}
    \item Sum-Synthesize: We train a ranker with the synthesized label and adopt the sum-based shallow tower for debiasing.
    \item PAL-Synthesize: We train a ranker with the synthesized label and adopt the PAL for debiasing.
    \item Sum-Click: We train the click and dwell-time task respectively and adopt sum-based shallow tower for debiasing on the click task.
    \item PAL-Click: We train the click and dwell-time task respectively and adopt PAL for debiasing on the click task.
    \item PAL-Click + PAL-DwellTime: We train the click and dwell-time task respectively and adopt PAL on both tasks.
\end{itemize}


For baseline models with multiple tasks, we combine the predicted click probability and dwell-time with the same weighted summation function as the  synthesized label for ranking. As shown in \autoref{table1}, our method outperforms all baseline methods significantly in both datasets. One can see that LambdaRank performs worse than pointwise DNN with the biased synthesized label, which indicates that debiasing is more important for pairwise learning in recommendation scenarios and can explain why pointwise methods are the mainstream rankers for recommender systems before. Moreover, employing shallow tower on dwell-time harms the performance significantly, which shows that shallow tower cannot work well for continuous labels. In contrast, our probabilistic graphical based method has a clear separation of bias and relevance and achieves superior performance on debiasing and preference learning for different  applications. Moreover, direct optimization for ranking metrics achieves significant improvement 
in both datasets. Besides, the pairwise trust bias correction exhibits better performance on Yahoo Dataset but similar performance on MovieLens. To explore why the pairwise trust bias correction performs differently on Yahoo Dataset and MovieLens, we analyze the statistics of $m_{ij}$ defined in Equation  \ref{eq:14} in these datasets and find that $m_{ij}$ in different position pairs are roughly similar in MovieLens but differs a lot in Yahoo Dataset, which indicates that there is no improvement room for pairwise trust bias correction in MovieLens.

\begin{figure*}[htbp]
\setlength{\abovecaptionskip}{-0.05cm} 
\centering  
\subfigure[Continuous Label]{ 
\begin{minipage}[t]{0.4\linewidth}
\includegraphics[scale=0.31]{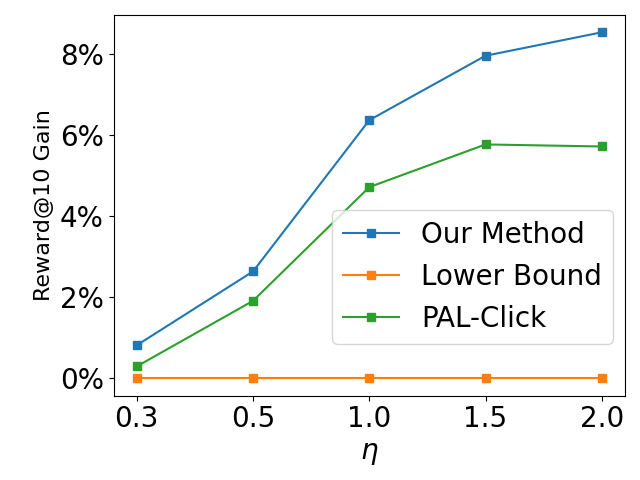}
\setlength{\abovecaptionskip}{-0.05cm} 
\end{minipage}
}
\subfigure[Categorical Label]{   
\begin{minipage}[t]{0.4\linewidth}\textbf{}
\includegraphics[scale=0.31]{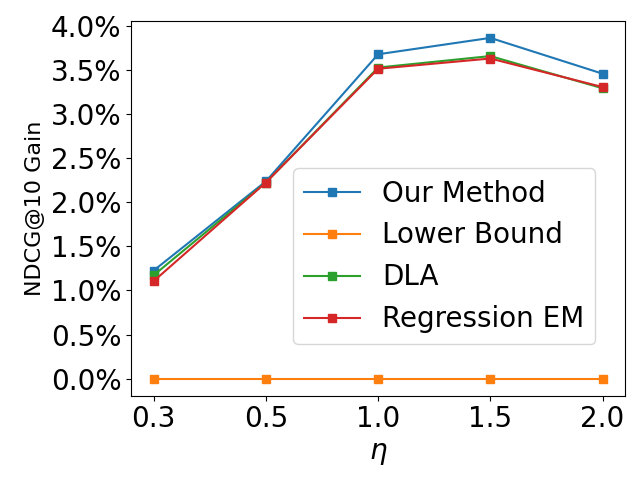}  
\setlength{\abovecaptionskip}{-0.05cm} 
\end{minipage}
}
\caption{Performance Gain over Lower Bound with Different Severity of Position Bias}    
\label{fig3}    
\end{figure*}
\vspace{-0.05cm}


\subsection{Experiments on Categorical Labels (RQ2)} \label{subsection54}
We also perform experiments on the sampled binary click labels of Yahoo Dataset and Web30K  to compare our method with the SOTA unbiased LTR methods including DLA \cite{ai2018unbiased}, Regression EM \cite{wang2018position} and Pairwise Debiasing \cite{hu2019unbiased} to answer RQ2. For a comprehensive comparison, we train LambdaRank models with the unbiased relevance labels and the biased click labels without debiasing as upper bound and lower bound respectively. It is shown in \autoref{table2} that our method performs competitively with the SOTA unbiased LTR methods on categorical labels. Furthermore, the pairwise trust bias correction achieves consistent improvement on both Yahoo Dataset and Web30K. Besides, direct optimization for ranking metrics exhibits better performance on Web30K but similar performance on Yahoo Dataset as we can only compute $\Delta Z_{ij}$ based on the noisy click signal. On the whole, our method achieves competitive performance on both continuous and categorical labels.

\subsection{Robustness Analysis (RQ3)}
To evaluate the robustness of the proposed method, we compare it with the best baseline methods, i.e. DLA, Regression EM and PAL-Click,  on Yahoo Dataset under different severity of position bias  and different weights of continuous labels in this section.

\subsubsection{Severity of Position Bias.}
\autoref{fig3} shows the performance gain of debiasing methods over the lower bound (LambdaRank)  on continuous and categorical labels with different severity of position bias controlled by $\eta$. The performance gain is defined as the relative improvement on the evaluation metric over lower bound. One can see that our method is robust to the severity of position bias and outperforms other competitors consistently under different $\eta$.


\subsubsection{Relative Weight of Continuous Label.} 
\begin{figure}[!htbp]
\setlength{\abovecaptionskip}{-0.1cm} 
	\centering\textbf{}
	\includegraphics[scale=0.3]{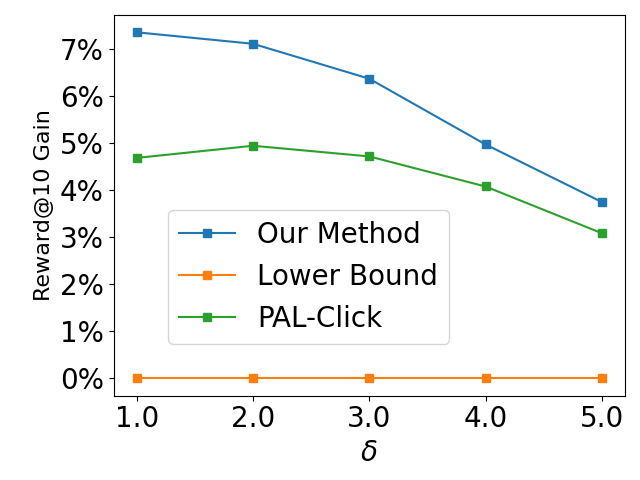}
	\caption{Performance Gain over Lower Bound with Different Relative Weights of Continuous Label}
	\label{fig4}
\end{figure}

According to \autoref{fig4}, our method performs best consistently under different weights of continuous labels controlled by $\delta$. On the whole, our method is robust to different severity of position bias and different weights of continuous labels.


\vspace{-0.3cm}
\vspace{-0.3cm}

\subsection{Online A/B Testing (RQ4)}
To evaluate the performance of the proposed method in real-world systems, careful online A/B testing has been conducted in the ranking module of a large-scale video recommender system at Tencent News for 7 days. The live metric and synthesized label are defined as the combination of multiple user actions (e.g. click, dwell-time, liking, and sharing) in the system. As PAL-Click is the best performed baseline method on continuous labels, we train multiple user actions respectively and employ PAL on the click task for debiasing as the baseline in online A/B testing. We implement all models in our deep learning framework, randomly distribute users into several buckets, and deploy each model to one of the buckets. Compared with PAL-Click, our method improves the live metric by 2.08\%, which demonstrates the effectiveness of our method in debiasing and preference learning in real-world applications.

\begin{figure}[!htbp]
\setlength{\abovecaptionskip}{-0.05cm} 
	\centering\textbf{}
	\includegraphics[scale=0.31]{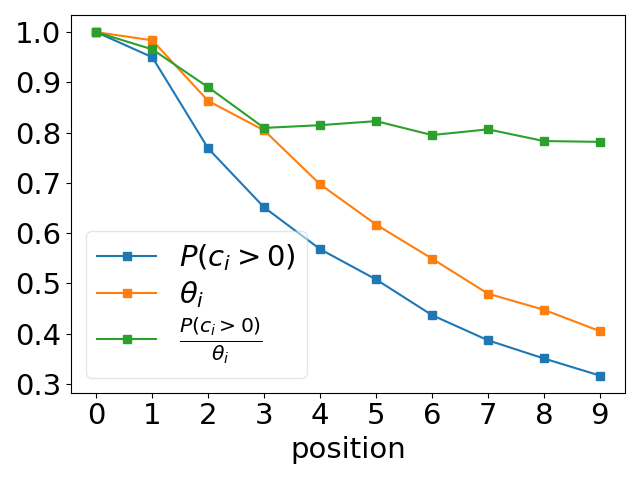}
	\caption{Statistical Probabilities and Learned Propensities}
	\label{fig5}
\end{figure}

To further investigate the performance of our  method in online A/B testing, we analyze the learned propensity per position denoted as $\theta_i$ of our method and compare it with the statistical probability of positive label in each position denoted as $P(c_i>0)$ in the system. \autoref{fig5} shows that $\theta_i$ is smaller for lower position, which is consistent with the decreasing trend of  $P(c_i>0)$. 
We also analyze the estimated probability of relevance at each position denoted as $\frac{P(c_i>0)}{\theta_i}$ according to Equation \ref{eq:3}. It is shown that $\frac{P(c_i>0)}{\theta_i}$ also exhibits a stepwise decreasing trend especially in higher positions, which is also consistent with the recommendation mechanism that more relevant items is usually exposed in higher-ranked positions. In summary, 
our method can learn and correct position bias effectively in real-world recommendation applications.

\vspace{-0.2cm}
\vspace{-0.3cm}

\section{Conclusion}
In this paper, we propose a novel unbiased pairwise LTR method to model position-based examination bias and trust bias in the pairwise fashion for better bias correction and preference learning, which is the first unbiased LTR method working for both categorical and continuous labels. Experiment results on public benchmark datasets and online  A/B testing show that the proposed method achieves significant improvements over SOTA ranking models for continuous labels and competitive performance with SOTA unbiased LTR methods for categorical labels. Correcting for other types of bias besides position bias will be the focus of future work.

\bibliographystyle{ACM-Reference-Format}
\balance
\bibliography{sigir}


\begin{thebibliography}{31}


\ifx \showCODEN    \undefined \def \showCODEN     #1{\unskip}     \fi
\ifx \showDOI      \undefined \def \showDOI       #1{#1}\fi
\ifx \showISBNx    \undefined \def \showISBNx     #1{\unskip}     \fi
\ifx \showISBNxiii \undefined \def \showISBNxiii  #1{\unskip}     \fi
\ifx \showISSN     \undefined \def \showISSN      #1{\unskip}     \fi
\ifx \showLCCN     \undefined \def \showLCCN      #1{\unskip}     \fi
\ifx \shownote     \undefined \def \shownote      #1{#1}          \fi
\ifx \showarticletitle \undefined \def \showarticletitle #1{#1}   \fi
\ifx \showURL      \undefined \def \showURL       {\relax}        \fi
\providecommand\bibfield[2]{#2}
\providecommand\bibinfo[2]{#2}
\providecommand\natexlab[1]{#1}
\providecommand\showeprint[2][]{arXiv:#2}

\bibitem[\protect\citeauthoryear{Agarwal, Wang, Li, Bendersky, and
  Najork}{Agarwal et~al\mbox{.}}{2019}]%
        {agarwal2019addressing}
\bibfield{author}{\bibinfo{person}{Aman Agarwal}, \bibinfo{person}{Xuanhui
  Wang}, \bibinfo{person}{Cheng Li}, \bibinfo{person}{Michael Bendersky}, {and}
  \bibinfo{person}{Marc Najork}.} \bibinfo{year}{2019}\natexlab{}.
\newblock \showarticletitle{Addressing trust bias for unbiased
  learning-to-rank}. In \bibinfo{booktitle}{\emph{The World Wide Web
  Conference}}. \bibinfo{pages}{4--14}.
\newblock


\bibitem[\protect\citeauthoryear{Ai, Bi, Luo, Guo, and Croft}{Ai
  et~al\mbox{.}}{2018}]%
        {ai2018unbiased}
\bibfield{author}{\bibinfo{person}{Qingyao Ai}, \bibinfo{person}{Keping Bi},
  \bibinfo{person}{Cheng Luo}, \bibinfo{person}{Jiafeng Guo}, {and}
  \bibinfo{person}{W~Bruce Croft}.} \bibinfo{year}{2018}\natexlab{}.
\newblock \showarticletitle{Unbiased learning to rank with unbiased propensity
  estimation}. In \bibinfo{booktitle}{\emph{The 41st International ACM SIGIR
  Conference on Research \& Development in Information Retrieval}}.
  \bibinfo{pages}{385--394}.
\newblock


\bibitem[\protect\citeauthoryear{Burges, Ragno, and Le}{Burges
  et~al\mbox{.}}{2006}]%
        {burges2006learning}
\bibfield{author}{\bibinfo{person}{Christopher Burges}, \bibinfo{person}{Robert
  Ragno}, {and} \bibinfo{person}{Quoc Le}.} \bibinfo{year}{2006}\natexlab{}.
\newblock \showarticletitle{Learning to rank with nonsmooth cost functions}.
\newblock \bibinfo{journal}{\emph{Advances in neural information processing
  systems}}  \bibinfo{volume}{19} (\bibinfo{year}{2006}),
  \bibinfo{pages}{193--200}.
\newblock


\bibitem[\protect\citeauthoryear{Burges, Shaked, Renshaw, Lazier, Deeds,
  Hamilton, and Hullender}{Burges et~al\mbox{.}}{2005}]%
        {burges2005learning}
\bibfield{author}{\bibinfo{person}{Chris Burges}, \bibinfo{person}{Tal Shaked},
  \bibinfo{person}{Erin Renshaw}, \bibinfo{person}{Ari Lazier},
  \bibinfo{person}{Matt Deeds}, \bibinfo{person}{Nicole Hamilton}, {and}
  \bibinfo{person}{Greg Hullender}.} \bibinfo{year}{2005}\natexlab{}.
\newblock \showarticletitle{Learning to rank using gradient descent}. In
  \bibinfo{booktitle}{\emph{Proceedings of the 22nd international conference on
  Machine learning}}. \bibinfo{pages}{89--96}.
\newblock


\bibitem[\protect\citeauthoryear{Burges}{Burges}{2010}]%
        {burges2010ranknet}
\bibfield{author}{\bibinfo{person}{Christopher~JC Burges}.}
  \bibinfo{year}{2010}\natexlab{}.
\newblock \showarticletitle{From ranknet to lambdarank to lambdamart: An
  overview}.
\newblock \bibinfo{journal}{\emph{Learning}} \bibinfo{volume}{11},
  \bibinfo{number}{23-581} (\bibinfo{year}{2010}), \bibinfo{pages}{81}.
\newblock


\bibitem[\protect\citeauthoryear{Capp{\'e} and Moulines}{Capp{\'e} and
  Moulines}{2009}]%
        {cappe2009line}
\bibfield{author}{\bibinfo{person}{Olivier Capp{\'e}} {and}
  \bibinfo{person}{Eric Moulines}.} \bibinfo{year}{2009}\natexlab{}.
\newblock \showarticletitle{On-line expectation--maximization algorithm for
  latent data models}.
\newblock \bibinfo{journal}{\emph{Journal of the Royal Statistical Society:
  Series B (Statistical Methodology)}} \bibinfo{volume}{71},
  \bibinfo{number}{3} (\bibinfo{year}{2009}), \bibinfo{pages}{593--613}.
\newblock


\bibitem[\protect\citeauthoryear{Chen, Dong, Wang, Feng, Wang, and He}{Chen
  et~al\mbox{.}}{2020}]%
        {chen2020bias}
\bibfield{author}{\bibinfo{person}{Jiawei Chen}, \bibinfo{person}{Hande Dong},
  \bibinfo{person}{Xiang Wang}, \bibinfo{person}{Fuli Feng},
  \bibinfo{person}{Meng Wang}, {and} \bibinfo{person}{Xiangnan He}.}
  \bibinfo{year}{2020}\natexlab{}.
\newblock \showarticletitle{Bias and debias in recommender system: A survey and
  future directions}.
\newblock \bibinfo{journal}{\emph{arXiv preprint arXiv:2010.03240}}
  (\bibinfo{year}{2020}).
\newblock


\bibitem[\protect\citeauthoryear{Guo, Yu, Liu, Tang, and Zhang}{Guo
  et~al\mbox{.}}{2019}]%
        {guo2019pal}
\bibfield{author}{\bibinfo{person}{Huifeng Guo}, \bibinfo{person}{Jinkai Yu},
  \bibinfo{person}{Qing Liu}, \bibinfo{person}{Ruiming Tang}, {and}
  \bibinfo{person}{Yuzhou Zhang}.} \bibinfo{year}{2019}\natexlab{}.
\newblock \showarticletitle{PAL: a position-bias aware learning framework for
  CTR prediction in live recommender systems}. In
  \bibinfo{booktitle}{\emph{Proceedings of the 13th ACM Conference on
  Recommender Systems}}. \bibinfo{pages}{452--456}.
\newblock


\bibitem[\protect\citeauthoryear{Hu, Koren, and Volinsky}{Hu
  et~al\mbox{.}}{2008}]%
        {hu2008collaborative}
\bibfield{author}{\bibinfo{person}{Yifan Hu}, \bibinfo{person}{Yehuda Koren},
  {and} \bibinfo{person}{Chris Volinsky}.} \bibinfo{year}{2008}\natexlab{}.
\newblock \showarticletitle{Collaborative filtering for implicit feedback
  datasets}. In \bibinfo{booktitle}{\emph{2008 Eighth IEEE International
  Conference on Data Mining}}. Ieee, \bibinfo{pages}{263--272}.
\newblock


\bibitem[\protect\citeauthoryear{Hu, Wang, Peng, and Li}{Hu
  et~al\mbox{.}}{2019}]%
        {hu2019unbiased}
\bibfield{author}{\bibinfo{person}{Ziniu Hu}, \bibinfo{person}{Yang Wang},
  \bibinfo{person}{Qu Peng}, {and} \bibinfo{person}{Hang Li}.}
  \bibinfo{year}{2019}\natexlab{}.
\newblock \showarticletitle{Unbiased LambdaMART: An unbiased pairwise
  learning-to-rank algorithm}. In \bibinfo{booktitle}{\emph{The World Wide Web
  Conference}}. \bibinfo{pages}{2830--2836}.
\newblock


\bibitem[\protect\citeauthoryear{Huang, Sharma, Sun, Xia, Zhang, Pronin,
  Padmanabhan, Ottaviano, and Yang}{Huang et~al\mbox{.}}{2020}]%
        {huang2020embedding}
\bibfield{author}{\bibinfo{person}{Jui-Ting Huang}, \bibinfo{person}{Ashish
  Sharma}, \bibinfo{person}{Shuying Sun}, \bibinfo{person}{Li Xia},
  \bibinfo{person}{David Zhang}, \bibinfo{person}{Philip Pronin},
  \bibinfo{person}{Janani Padmanabhan}, \bibinfo{person}{Giuseppe Ottaviano},
  {and} \bibinfo{person}{Linjun Yang}.} \bibinfo{year}{2020}\natexlab{}.
\newblock \showarticletitle{Embedding-based retrieval in facebook search}. In
  \bibinfo{booktitle}{\emph{Proceedings of the 26th ACM SIGKDD International
  Conference on Knowledge Discovery \& Data Mining}}.
  \bibinfo{pages}{2553--2561}.
\newblock


\bibitem[\protect\citeauthoryear{J{\"a}rvelin and
  Kek{\"a}l{\"a}inen}{J{\"a}rvelin and Kek{\"a}l{\"a}inen}{2002}]%
        {jarvelin2002cumulated}
\bibfield{author}{\bibinfo{person}{Kalervo J{\"a}rvelin} {and}
  \bibinfo{person}{Jaana Kek{\"a}l{\"a}inen}.} \bibinfo{year}{2002}\natexlab{}.
\newblock \showarticletitle{Cumulated gain-based evaluation of IR techniques}.
\newblock \bibinfo{journal}{\emph{ACM Transactions on Information Systems
  (TOIS)}} \bibinfo{volume}{20}, \bibinfo{number}{4} (\bibinfo{year}{2002}),
  \bibinfo{pages}{422--446}.
\newblock


\bibitem[\protect\citeauthoryear{J{\"a}rvelin and
  Kek{\"a}l{\"a}inen}{J{\"a}rvelin and Kek{\"a}l{\"a}inen}{2017}]%
        {jarvelin2017ir}
\bibfield{author}{\bibinfo{person}{Kalervo J{\"a}rvelin} {and}
  \bibinfo{person}{Jaana Kek{\"a}l{\"a}inen}.} \bibinfo{year}{2017}\natexlab{}.
\newblock \showarticletitle{IR evaluation methods for retrieving highly
  relevant documents}. In \bibinfo{booktitle}{\emph{ACM SIGIR Forum}},
  Vol.~\bibinfo{volume}{51}. ACM New York, NY, USA, \bibinfo{pages}{243--250}.
\newblock


\bibitem[\protect\citeauthoryear{Joachims, Granka, Pan, Hembrooke, and
  Gay}{Joachims et~al\mbox{.}}{2017a}]%
        {joachims2017accurately}
\bibfield{author}{\bibinfo{person}{Thorsten Joachims}, \bibinfo{person}{Laura
  Granka}, \bibinfo{person}{Bing Pan}, \bibinfo{person}{Helene Hembrooke},
  {and} \bibinfo{person}{Geri Gay}.} \bibinfo{year}{2017}\natexlab{a}.
\newblock \showarticletitle{Accurately interpreting clickthrough data as
  implicit feedback}. In \bibinfo{booktitle}{\emph{ACM SIGIR Forum}},
  Vol.~\bibinfo{volume}{51}. Acm New York, NY, USA, \bibinfo{pages}{4--11}.
\newblock


\bibitem[\protect\citeauthoryear{Joachims, Swaminathan, and Schnabel}{Joachims
  et~al\mbox{.}}{2017b}]%
        {joachims2017unbiased}
\bibfield{author}{\bibinfo{person}{Thorsten Joachims}, \bibinfo{person}{Adith
  Swaminathan}, {and} \bibinfo{person}{Tobias Schnabel}.}
  \bibinfo{year}{2017}\natexlab{b}.
\newblock \showarticletitle{Unbiased learning-to-rank with biased feedback}. In
  \bibinfo{booktitle}{\emph{Proceedings of the Tenth ACM International
  Conference on Web Search and Data Mining}}. \bibinfo{pages}{781--789}.
\newblock


\bibitem[\protect\citeauthoryear{Johnson}{Johnson}{2014}]%
        {johnson2014logistic}
\bibfield{author}{\bibinfo{person}{Christopher~C Johnson}.}
  \bibinfo{year}{2014}\natexlab{}.
\newblock \showarticletitle{Logistic matrix factorization for implicit feedback
  data}.
\newblock \bibinfo{journal}{\emph{Advances in Neural Information Processing
  Systems}} \bibinfo{volume}{27}, \bibinfo{number}{78} (\bibinfo{year}{2014}),
  \bibinfo{pages}{1--9}.
\newblock


\bibitem[\protect\citeauthoryear{Koren, Bell, and Volinsky}{Koren
  et~al\mbox{.}}{2009}]%
        {koren2009matrix}
\bibfield{author}{\bibinfo{person}{Yehuda Koren}, \bibinfo{person}{Robert
  Bell}, {and} \bibinfo{person}{Chris Volinsky}.}
  \bibinfo{year}{2009}\natexlab{}.
\newblock \showarticletitle{Matrix factorization techniques for recommender
  systems}.
\newblock \bibinfo{journal}{\emph{Computer}} \bibinfo{volume}{42},
  \bibinfo{number}{8} (\bibinfo{year}{2009}), \bibinfo{pages}{30--37}.
\newblock


\bibitem[\protect\citeauthoryear{Lee, Park, and Lee}{Lee et~al\mbox{.}}{2021}]%
        {lee2021dual}
\bibfield{author}{\bibinfo{person}{Jae-woong Lee}, \bibinfo{person}{Seongmin
  Park}, {and} \bibinfo{person}{Jongwuk Lee}.} \bibinfo{year}{2021}\natexlab{}.
\newblock \showarticletitle{Dual Unbiased Recommender Learning for Implicit
  Feedback}. In \bibinfo{booktitle}{\emph{Proceedings of the 44th International
  ACM SIGIR Conference on Research and Development in Information Retrieval}}.
  \bibinfo{pages}{1647--1651}.
\newblock


\bibitem[\protect\citeauthoryear{Liang, Charlin, McInerney, and Blei}{Liang
  et~al\mbox{.}}{2016}]%
        {liang2016modeling}
\bibfield{author}{\bibinfo{person}{Dawen Liang}, \bibinfo{person}{Laurent
  Charlin}, \bibinfo{person}{James McInerney}, {and} \bibinfo{person}{David~M
  Blei}.} \bibinfo{year}{2016}\natexlab{}.
\newblock \showarticletitle{Modeling user exposure in recommendation}. In
  \bibinfo{booktitle}{\emph{Proceedings of the 25th international conference on
  World Wide Web}}. \bibinfo{pages}{951--961}.
\newblock


\bibitem[\protect\citeauthoryear{Rendle, Freudenthaler, Gantner, and
  Schmidt-Thieme}{Rendle et~al\mbox{.}}{2012}]%
        {rendle2012bpr}
\bibfield{author}{\bibinfo{person}{Steffen Rendle}, \bibinfo{person}{Christoph
  Freudenthaler}, \bibinfo{person}{Zeno Gantner}, {and} \bibinfo{person}{Lars
  Schmidt-Thieme}.} \bibinfo{year}{2012}\natexlab{}.
\newblock \showarticletitle{BPR: Bayesian personalized ranking from implicit
  feedback}.
\newblock \bibinfo{journal}{\emph{arXiv preprint arXiv:1205.2618}}
  (\bibinfo{year}{2012}).
\newblock


\bibitem[\protect\citeauthoryear{Reynolds}{Reynolds}{2009}]%
        {reynolds2009gaussian}
\bibfield{author}{\bibinfo{person}{Douglas~A Reynolds}.}
  \bibinfo{year}{2009}\natexlab{}.
\newblock \showarticletitle{Gaussian mixture models.}
\newblock \bibinfo{journal}{\emph{Encyclopedia of biometrics}}
  \bibinfo{volume}{741}, \bibinfo{number}{659-663} (\bibinfo{year}{2009}).
\newblock


\bibitem[\protect\citeauthoryear{Richardson, Dominowska, and Ragno}{Richardson
  et~al\mbox{.}}{2007}]%
        {richardson2007predicting}
\bibfield{author}{\bibinfo{person}{Matthew Richardson}, \bibinfo{person}{Ewa
  Dominowska}, {and} \bibinfo{person}{Robert Ragno}.}
  \bibinfo{year}{2007}\natexlab{}.
\newblock \showarticletitle{Predicting clicks: estimating the click-through
  rate for new ads}. In \bibinfo{booktitle}{\emph{Proceedings of the 16th
  international conference on World Wide Web}}. \bibinfo{pages}{521--530}.
\newblock


\bibitem[\protect\citeauthoryear{Saito}{Saito}{2020}]%
        {saito2020unbiased2}
\bibfield{author}{\bibinfo{person}{Yuta Saito}.}
  \bibinfo{year}{2020}\natexlab{}.
\newblock \showarticletitle{Unbiased Pairwise Learning from Biased Implicit
  Feedback}. In \bibinfo{booktitle}{\emph{Proceedings of the 2020 ACM SIGIR on
  International Conference on Theory of Information Retrieval}}.
  \bibinfo{pages}{5--12}.
\newblock


\bibitem[\protect\citeauthoryear{Saito, Yaginuma, Nishino, Sakata, and
  Nakata}{Saito et~al\mbox{.}}{2020}]%
        {saito2020unbiased}
\bibfield{author}{\bibinfo{person}{Yuta Saito}, \bibinfo{person}{Suguru
  Yaginuma}, \bibinfo{person}{Yuta Nishino}, \bibinfo{person}{Hayato Sakata},
  {and} \bibinfo{person}{Kazuhide Nakata}.} \bibinfo{year}{2020}\natexlab{}.
\newblock \showarticletitle{Unbiased recommender learning from
  missing-not-at-random implicit feedback}. In
  \bibinfo{booktitle}{\emph{Proceedings of the 13th International Conference on
  Web Search and Data Mining}}. \bibinfo{pages}{501--509}.
\newblock


\bibitem[\protect\citeauthoryear{Tang, Liu, Zhao, and Gong}{Tang
  et~al\mbox{.}}{2020}]%
        {tang2020progressive}
\bibfield{author}{\bibinfo{person}{Hongyan Tang}, \bibinfo{person}{Junning
  Liu}, \bibinfo{person}{Ming Zhao}, {and} \bibinfo{person}{Xudong Gong}.}
  \bibinfo{year}{2020}\natexlab{}.
\newblock \showarticletitle{Progressive layered extraction (ple): A novel
  multi-task learning (mtl) model for personalized recommendations}. In
  \bibinfo{booktitle}{\emph{Fourteenth ACM Conference on Recommender Systems}}.
  \bibinfo{pages}{269--278}.
\newblock


\bibitem[\protect\citeauthoryear{Wang, Bendersky, Metzler, and Najork}{Wang
  et~al\mbox{.}}{2016}]%
        {wang2016learning}
\bibfield{author}{\bibinfo{person}{Xuanhui Wang}, \bibinfo{person}{Michael
  Bendersky}, \bibinfo{person}{Donald Metzler}, {and} \bibinfo{person}{Marc
  Najork}.} \bibinfo{year}{2016}\natexlab{}.
\newblock \showarticletitle{Learning to rank with selection bias in personal
  search}. In \bibinfo{booktitle}{\emph{Proceedings of the 39th International
  ACM SIGIR conference on Research and Development in Information Retrieval}}.
  \bibinfo{pages}{115--124}.
\newblock


\bibitem[\protect\citeauthoryear{Wang, Golbandi, Bendersky, Metzler, and
  Najork}{Wang et~al\mbox{.}}{2018a}]%
        {wang2018position}
\bibfield{author}{\bibinfo{person}{Xuanhui Wang}, \bibinfo{person}{Nadav
  Golbandi}, \bibinfo{person}{Michael Bendersky}, \bibinfo{person}{Donald
  Metzler}, {and} \bibinfo{person}{Marc Najork}.}
  \bibinfo{year}{2018}\natexlab{a}.
\newblock \showarticletitle{Position bias estimation for unbiased learning to
  rank in personal search}. In \bibinfo{booktitle}{\emph{Proceedings of the
  Eleventh ACM International Conference on Web Search and Data Mining}}.
  \bibinfo{pages}{610--618}.
\newblock


\bibitem[\protect\citeauthoryear{Wang, Li, Golbandi, Bendersky, and
  Najork}{Wang et~al\mbox{.}}{2018b}]%
        {wang2018lambdaloss}
\bibfield{author}{\bibinfo{person}{Xuanhui Wang}, \bibinfo{person}{Cheng Li},
  \bibinfo{person}{Nadav Golbandi}, \bibinfo{person}{Michael Bendersky}, {and}
  \bibinfo{person}{Marc Najork}.} \bibinfo{year}{2018}\natexlab{b}.
\newblock \showarticletitle{The lambdaloss framework for ranking metric
  optimization}. In \bibinfo{booktitle}{\emph{Proceedings of the 27th ACM
  International Conference on Information and Knowledge Management}}.
  \bibinfo{pages}{1313--1322}.
\newblock


\bibitem[\protect\citeauthoryear{Wu, Chen, Zhao, He, Yin, and Chang}{Wu
  et~al\mbox{.}}{2021}]%
        {wu2021unbiased}
\bibfield{author}{\bibinfo{person}{Xinwei Wu}, \bibinfo{person}{Hechang Chen},
  \bibinfo{person}{Jiashu Zhao}, \bibinfo{person}{Li He},
  \bibinfo{person}{Dawei Yin}, {and} \bibinfo{person}{Yi Chang}.}
  \bibinfo{year}{2021}\natexlab{}.
\newblock \showarticletitle{Unbiased learning to rank in feeds recommendation}.
  In \bibinfo{booktitle}{\emph{Proceedings of the 14th ACM International
  Conference on Web Search and Data Mining}}. \bibinfo{pages}{490--498}.
\newblock


\bibitem[\protect\citeauthoryear{Yin, Luo, Lee, and Wang}{Yin
  et~al\mbox{.}}{2013}]%
        {yin2013silence}
\bibfield{author}{\bibinfo{person}{Peifeng Yin}, \bibinfo{person}{Ping Luo},
  \bibinfo{person}{Wang-Chien Lee}, {and} \bibinfo{person}{Min Wang}.}
  \bibinfo{year}{2013}\natexlab{}.
\newblock \showarticletitle{Silence is also evidence: interpreting dwell time
  for recommendation from psychological perspective}. In
  \bibinfo{booktitle}{\emph{Proceedings of the 19th ACM SIGKDD international
  conference on Knowledge discovery and data mining}}.
  \bibinfo{pages}{989--997}.
\newblock


\bibitem[\protect\citeauthoryear{Zhao, Hong, Wei, Chen, Nath, Andrews,
  Kumthekar, Sathiamoorthy, Yi, and Chi}{Zhao et~al\mbox{.}}{2019}]%
        {zhao2019recommending}
\bibfield{author}{\bibinfo{person}{Zhe Zhao}, \bibinfo{person}{Lichan Hong},
  \bibinfo{person}{Li Wei}, \bibinfo{person}{Jilin Chen},
  \bibinfo{person}{Aniruddh Nath}, \bibinfo{person}{Shawn Andrews},
  \bibinfo{person}{Aditee Kumthekar}, \bibinfo{person}{Maheswaran
  Sathiamoorthy}, \bibinfo{person}{Xinyang Yi}, {and} \bibinfo{person}{Ed
  Chi}.} \bibinfo{year}{2019}\natexlab{}.
\newblock \showarticletitle{Recommending what video to watch next: a multitask
  ranking system}. In \bibinfo{booktitle}{\emph{Proceedings of the 13th ACM
  Conference on Recommender Systems}}. \bibinfo{pages}{43--51}.
\newblock


\end{thebibliography}

\appendix

\end{document}


\maketitle




\begin{abstract}
Here, we provide a supplementary material for the paper ``Unbiased Pairwise Learning to Rank in Recommender Systems''. In detail, we illustrate the proofs of important formulas in the paper in detail.
\end{abstract}

\section{Proofs of Unbiased LTR for Continuous Labels}
In this section, we explain the derivation of the Equation 9 in the paper more specifically as follows:
\begin{align}
&  R_{unbiased}(f) \nonumber  \\
&=  \int_{}{}  \frac {L(\hat{y}_i,  c_i, \hat{y}_j,  c_j)} {\theta_i \theta_j } P(  e_j   =   1|u, i,  j,  x_i,  x_j, c_i>c_j  ) dP( c_i  >  c_j,  u,  x_i,  x_j,  i , j )   \\
&=  \int_{}{}  \frac {L(\hat{y}_i,  c_i, \hat{y}_j,  c_j)} {\theta_i \theta_j }  dP( c_i  >  c_j,  u,  x_i,  x_j,  i , j ) P(  e_j   =   1|u, i,  j,  x_i,  x_j, c_i>c_j  )   \\
& =  \int_{}{}  \frac {L(\hat{y}_i,  c_i,  \hat{y}_i,  c_j)} {\theta_i \theta_j }  dP( c_i  >  c_j,  u,  x_i,  x_j,i,j ,e_i=1,e_j=1)    \\
&=  \int_{}{}  \frac {L(\hat{y}_i,  c_i,  \hat{y}_i,  c_j)} {\theta_i \theta_j }  dP( r_i  >  r_j,  u,  x_i,  x_j,i,j ,e_i=1,e_j=1)     \\
&=  \int_{}{}  \frac {L(\hat{y}_i,  r_i,  \hat{y}_j,  r_j)} {\theta_i \theta_j }  dP( r_i \! > \! r_j,  u,  x_i,  x_j, i,j) P(e_i \! = \! 1|r_i \! > \! r_j,  u,  x_i,  x_j,i,j)P(e_j \!= \!  1|r_i \! > \!  r_j,  u,  x_i,  x_j, i,j,e_i \!= \!1)    \\
&=  \int_{}{}  \frac {L(\hat{y}_i,  r_i,  \hat{y}_j,  r_j)} {\theta_i \theta_j }  dP( r_i  >  r_j,  u,  x_i,  x_j,i,j) P(e_i  =  1|i)P(e_j =  1|j)    \\
&=  \int_{}{}  {L(\hat{y}_i,  r_i,  \hat{y}_i,  r_j)} dP( r_i  >  r_j,  u,  x_i,  x_j,i,j) \\
&=  \int_{}{}  {L(\hat{y}_i,  r_i,  \hat{y}_i,  r_j)} dP( r_i  >  r_j,  u,  x_i,  x_j )  = R_{rel}(f)
\end{align}
Specifically, we can derive the integration in Equation 1 to Equation 2 through the method of integration by substitution. Equation 3 is based on the Bayes law. According to the assumption that $c_i>c_j$ if and only if $r_i>r_j$ and $x_i$ and $x_j$ are both examined, we can derive the Equation 4. Similarly, Equation 5 is also based on the Bayes law  and Equation 6 is based on the assumption that the examination of different positions and the relative relevance between items are mutually independent. We can than derive Equation 7 by the  definition of $\theta_i$ and $\theta_j$. Finally, $R_{ref}(f)$ can be derived as loss function is independent of exposed positions.

\section{Proofs of Formulas in Expectation Step}
In this section, we derive all the formulas of Expectation step shown in Figure 1 in the paper. First, the first equation in Figure 1 can be derived as follows.
\begin{align}
& P(e_i=1,e_j=1,r_i>r_j|u,x_i,x_j,i,j,c_i>c_j) \nonumber \\
& = \frac{P(c_i>c_j | e_i=1,e_j=1,r_i>r_j,u,x_i,x_j,i,j)*P(e_i=1,e_j=1,r_i>r_j|u,x_i,x_j,i,j)}{P(c_i>c_j|u,x_i,x_j,i,j)} \nonumber \\
& = \frac{\epsilon_{ij}^+P(e_i=1|i) P(e_j=1|j)P(r_i>r_j|u,x_i,x_j)}{P(c_i>c_j|u,x_i,x_j,i,j)}  \nonumber \\
& = \frac { \theta_i \theta_j \epsilon_{ij}^+ \gamma_{u,x_i,x_j} } { \theta_i \theta_j  ( \epsilon_{ij}^+ \gamma_{u,x_i,x_j}  +  \epsilon_{ij}^-(1  -  \gamma_{u,x_i,x_j}))  +  \theta_i ( 1  -  \theta_j) \beta_{u,x_i}} 
\end{align}
The first step is based on Bayes law. The second step is based on the definition of $\epsilon_{ij}^+$ and the independent assumption between $e_i$ ,$e_j$ and $r_i>r_j$. Similarly, we can derive other formulas as follows:

\begin{align}
& P(e_i=1,e_j=1,r_i \leq r_j|u,x_i,x_j,i,j,c_i>c_j) \nonumber \\
& = \frac{P(c_i>c_j | e_i=1,e_j=1,r_i \leq r_j,u,x_i,x_j,i,j)*P(e_i=1,e_j=1,r_i \leq r_j|u,x_i,x_j,i,j)}{P(c_i>c_j|u,x_i,x_j,i,j)} \nonumber \\
& = \frac{\epsilon_{ij}^-P(e_i=1|i) P(e_j=1|j)P(r_i \leq r_j|u,x_i,x_j)}{P(c_i>c_j|u,x_i,x_j,i,j)} \nonumber \\
& = \frac { \theta_i \theta_j \epsilon_{ij}^- (1-\gamma_{u,x_i,x_j}) } 
{ \theta_i \theta_j  ( \epsilon_{ij}^+ \gamma_{u,x_i,x_j}  +  \epsilon_{ij}^-(1  -  \gamma_{u,x_i,x_j}))  +  \theta_i ( 1  -  \theta_j) \beta_{u,x_i}}  
\end{align}
\begin{align}
& P(e_i=1,e_j=0,r_i > 0|u,x_i,x_j,i,j,c_i>c_j) \nonumber \\
& = \frac{P(c_i>c_j | e_i=1,e_j=0,r_i > 0,u,x_i,x_j,i,j)*P(e_i=1,e_j=0,r_i  > 0|u,x_i,x_j,i,j)}{P(c_i>c_j|u,x_i,x_j,i,j)} \nonumber \\
& = \frac{P(c_i>0 | e_i=1,r_i > 0,u,x_i,i)*P(e_i=1,e_j=0,r_i  > 0|u,x_i,x_j,i,j)}{P(c_i>c_j|u,x_i,x_j,i,j)} \nonumber \\
& = \frac{1*P(e_i=1|i)P(e_j=0|j)P(r_i > 0|u, x_i)}{P(c_i>c_j|u,x_i,x_j,i,j)}  \nonumber \\
& = \frac { \theta_i (1-\theta_j) \beta_{u,x_i} } 
{ \theta_i \theta_j  ( \epsilon_{ij}^+ \gamma_{u,x_i,x_j}  +  \epsilon_{ij}^-(1  -  \gamma_{u,x_i,x_j}))  +  \theta_i ( 1  -  \theta_j) \beta_{u,x_i}}  
\end{align}
\begin{align}
& P(e_i=1,e_j=1,r_i \leq r_j|u,x_i,x_j,i,j,c_i \leq c_j) \nonumber \\
& = \frac{P(c_i \leq c_j | e_i=1,e_j=1,r_i \leq r_j,u,x_i,x_j,i,j)*P(e_i=1,e_j=1,r_i \leq r_j|u,x_i,x_j,i,j)}{P(c_i \leq c_j|u,x_i,x_j,i,j)} \nonumber \\
& = \frac{ (1-\epsilon_{ij}^-)*P(e_i=1|i)P(e_j=1|j)P(r_i \leq r_j|u,x_i,x_j)}{1-P(c_i > c_j|u,x_i,x_j,i,j)} \nonumber \\
& = \frac { \theta_i \theta_j (1-\epsilon_{ij}^-) (1-\gamma_{u,x_i,x_j}) }
{1- (\theta_i \theta_j  ( \epsilon_{ij}^+ \gamma_{u,x_i,x_j}  +  \epsilon_{ij}^-(1  -  \gamma_{u,x_i,x_j}))  +  \theta_i ( 1  -  \theta_j) \beta_{u,x_i})} 
\end{align}
\begin{align}
& P(e_i=1,e_j=1,r_i > r_j|u,x_i,x_j,i,j,c_i \leq c_j) \nonumber \\
& = \frac{P(c_i \leq c_j | e_i=1,e_j=1,r_i > r_j,u,x_i,x_j,i,j)*P(e_i=1,e_j=1,r_i > r_j|u,x_i,x_j,i,j)}{P(c_i \leq c_j|u,x_i,x_j,i,j)} \nonumber \\
& = \frac{ (1-\epsilon_{ij}^+)*P(e_i=1|i)P(e_j=1|j)P(r_i > r_j|u,x_i,x_j)}{1-P(c_i > c_j|u,x_i,x_j,i,j)} \nonumber \\
& = \frac { \theta_i \theta_j (1-\epsilon_{ij}^+) \gamma_{u,x_i,x_j} }
{1- (\theta_i \theta_j  ( \epsilon_{ij}^+ \gamma_{u,x_i,x_j}  +  \epsilon_{ij}^-(1  -  \gamma_{u,x_i,x_j}))  +  \theta_i ( 1  -  \theta_j) \beta_{u,x_i})}  \\
\nonumber \\
& P(e_i=0,r_i > 0|u,x_i,i,c_i=0) \nonumber \\
& = \frac{P(c_i=0|e_i=0,r_i>0,u,x_i,i)P(e_i=0,r_i>0|u,x_i,i)}{P(c_i=0|u,x_i,i)} \nonumber \\
& = \frac{1*P(e_i=0|i)P(r_i>0|u,x_i)}{P(c_i=0|u,x_i,i)} \nonumber \\
& =
 \frac{(1-\theta_i)\beta_{u,x_i}}
{1-\theta_i\beta_{u,x_i}}  \\
\nonumber \\
& P(e_i=1,r_i \leq 0|u,x_i,i,c_i=0)   \nonumber \\ 
& =
\frac{P(c_i=0|e_i=1,r_i \leq 0,u,x_i,i)P(e_i=1,r_i \leq 0|u,x_i,i)}{P(c_i=0|u,x_i,i)} \nonumber \\ 
& = \frac{1*P(e_i=1|i)P(r_i \leq 0|u,x_i)}{P(c_i=0|u,x_i,i)} \nonumber \\ 
& =
\frac{\theta_i(1-\beta_{u,x_i})}
{1-\theta_i\beta_{u,x_i}} 
\end{align}

